\documentclass[aps,prc,superscriptaddress,showpacs]{revtex4-1}	
 \usepackage{amsmath}  
 \usepackage{makeidx}
 \usepackage{graphicx,color}
 \usepackage{amsfonts}
 \usepackage[ansinew]{inputenc}
 \usepackage[usenames,dvipsnames]{pstricks}
 \usepackage{subfigure}
 \usepackage{epsfig}
 \usepackage{pst-grad}
 \usepackage{pst-plot}
 \usepackage{mathrsfs}
\usepackage[figuresright]{rotating}
 \def\vect#1{\mbox{\boldmath $#1$}}
	          \newcommand{\oo}{${{^{16}}{\rm O}}$}
              \newcommand{\nene}{${^{20}{\rm Ne}}$}
              \newcommand{\cc}{${^{12}{\rm C}}$}
              \makeindex            
\def\Vec#1{\mbox{\boldmath $#1$}}              
\begin{document}
\title{Nonlocalized cluster dynamics and nuclear molecular structure}
\author{Bo Zhou}
\email{zhoubo@rcnp.osaka-u.ac.jp.}
\affiliation{Department of Physics, Nanjing University, Nanjing 210093, China}
\affiliation{Research Center for Nuclear Physics (RCNP), Osaka University, Osaka 567-0047, Japan}
\affiliation{Nishina Center for Accelerator-Based Science, The institute of Physical and Chemical Research (RIKEN), Wako 351-0198, Japan}
\author{Yasuro Funaki}
\email{funaki@riken.jp.}
\affiliation{Nishina Center for Accelerator-Based Science, The institute of Physical and Chemical Research (RIKEN), Wako 351-0198, Japan}
 \author{Hisashi Horiuchi}
 \affiliation {Research Center for Nuclear Physics (RCNP), Osaka University, Osaka  567-0047, Japan}
 \affiliation {International Institute for Advanced Studies, Kizugawa 619-0225,  Japan}	
\author{Zhongzhou Ren} 
\email{zren@nju.edu.cn.}
\affiliation{Department of Physics, Nanjing University, Nanjing 210093, China}
\affiliation{Center of Theoretical Nuclear Physics, National Laboratory of Heavy-Ion Accelerator, Lanzhou 730000, China}  
\author{\mbox{Gerd R\"{o}pke}}  
\affiliation{Institut f\"{u}r Physik, Universit\"{a}t Rostock, D-18051 Rostock, Germany}
\author{Peter Schuck}
\affiliation{Institut de Physique Nucl\'{e}aire, Universit\'e Paris-Sud, IN2P3-CNRS, UMR 8608, F-91406, Orsay, France}
\affiliation{Laboratoire de Physique et Mod\'elisation des Milieux Condens\'es, CNRS-UMR 5493, F-38042 Grenoble Cedex 9, France}
 \author{Akihiro Tohsaki}
 \affiliation{Research Center for Nuclear Physics (RCNP), Osaka University, Osaka 567-0047, Japan}
\author{Chang Xu}
\affiliation{Department of Physics, Nanjing University, Nanjing 210093, China}
\author{Taiichi Yamada}
\affiliation{Laboratory of Physics, Kanto Gakuin University, Yokohama 236-8501, Japan}

\date{\today}

\begin{abstract}
A container picture is proposed for understanding cluster dynamics where the clusters make nonlocalized motion occupying the lowest orbit of the cluster mean-field potential characterized by the size parameter $``B"$ in the THSR (Tohsaki-Horiuchi-Schuck-R\"{o}pke) wave function. The nonlocalized cluster aspects of the inversion-doublet bands in \nene\ which have been considered as a typical manifestation of localized clustering are discussed. So far unexplained puzzling features of the THSR wave function, namely that after angular-momentum projection for two cluster systems the prolate THSR wave function is almost 100$\%$ equivalent to an oblate THSR wave function is clarified. It is shown that the true intrinsic two-cluster THSR configuration is nonetheless prolate. The proposal of the container picture is based on the fact that typical cluster systems, 2$\alpha$, 3$\alpha$, and $\alpha$+\oo, are all well described by a {\it single} THSR wave function. It will be shown for the case of linear-chain states with two and three $\alpha$-clusters  as well as for the $\alpha$+\oo\ system that localization is entirely of kinematical origin, that is, due to the inter-cluster Pauli repulsion. It is concluded that this feature is general for nuclear cluster states. 
\end{abstract}   
 
                                         
\maketitle

\section{Introduction  \label{intro}}
 The formation of clusters is a general problem in many-body physics which occurs in various systems. 
 In particular, it is one of the most important features in light nuclei~\cite{Wi77,Oe06,Fr07,Ho12} together with
 the formation of the usual nucleon mean field. A very novel cluster state in light nuclei is the $\alpha$-condensate-like state which has attracted increasing interest in recent 
 years~\cite{To01,Fu02,Fu03,Ya04,Ya05,Fu05,Fu06,tz05,tz06,enyo07,tk07,freer09,hyld10,ohku10,cher10,itoh11,wrz11,freer11,
 hak12,Fu08,Fu08a,Fu09,Fu10,Fu10a,Ya12,Ya12a,itagaki07,itagaki08}. This state can be considered as a gas-like state of clusters in which the center-of-mass motion of each $\alpha$ cluster in nuclei occupies the same $0s$ orbit. The THSR (Tohsaki-Horiuchi-Schuck-R\"{o}pke) wave function~\cite{To01} proposed for treating the $\alpha$-condensate-like state has been proved to be very suitable for the realistic description of the dilute gas-like state of clusters. Actually in the case of the Hoyle state (the second $0^+$ state)  of $^{12}$C, it was found that the full microscopic solutions~\cite{Ka81} of the 3$\alpha$ RGM (resonating group method)~\cite{Sup77}  and that~\cite{Ue77} of the GCM (generator coordinate method) with the Brink wave function~\cite{Sup77} are almost 
100$\%$ equivalent to single 3$\alpha$ THSR wave functions~\cite{Fu03}. Also in the case of the ground-state band of $^8$Be, the full microscopic solutions of the 2$\alpha$ RGM or its equivalent GCM with the Brink wave function were found to be practically 100$\%$ equivalent to single 2$\alpha$ THSR wave functions~\cite{Fu02}.  
 
Although the THSR wave function was devised for describing the gas-like state of clusters, it was already found 
a decade ago in Ref.~\cite{Fu03} that the wave functions of the ground state of $^{12}$C with normal density obtained by 
3$\alpha$ RGM and 3$\alpha$ Brink-GCM calculations have a large value of about 0.93 for the squared overlap with {\it single} 3$\alpha$ THSR wave functions.  Recently the present authors found~\cite{Zh12} that the $^{16}$O + $\alpha$ Brink-GCM wave functions of the states of the ground-state rotational band of $^{20}$Ne with normal density are almost 100$\%$ equivalent to single $^{16}$O + $\alpha$ THSR wave functions.   These results show that the THSR wave function has an ability which was not expected at first, namely it can be used to study not only the gas-like cluster states with low density but also (cluster) states  with normal density.  
 
 In $^{20}$Ne, the ground-state rotational band with $K^\pi = 0_1^+$ is known to constitute an inversion doublet together with the negative-parity rotational band with $K^\pi = 0_1^-$ built upon the $1^-$ state at the excitation energy $Ex$ = 5.79 MeV. The existence of the inversion doublet bands has been regarded as being a clear manifestation of the existence of 
 the parity-violating intrinsic state due to the $^{16}$O + $\alpha$ localized clustering.  In general, 
 in non-identical two-cluster systems, the existence of inversion-doublet bands  has been regarded as a clear 
 indication of the existence of the localized cluster structure together with the observation of the large cluster 
 decay widths.   This argument implies that we have to regard the states of the ground-state rotational band of $^{20}$Ne 
 as having a $^{16}$O + $\alpha$ localized clustering.  However, the THSR wave function is the wave function of 
 nonlocalized clusters as is shown by the fact that it was devised for describing cluster-gas-like states.  
 This puzzle of localized or nonlocalized clustering  which is found in the ground-state band states of $^{20}$Ne is the same as that presented in Ref.~\cite{Fu02}, where it was shown that the ground-state band states of 
 $^8$Be which has been regarded long since as having the localized $\alpha$ + $\alpha$ structure are very well described by single 2$\alpha$ THSR wave functions which are the wave functions of nonlocalized clusters. 
 
 In order to investigate in detail the above-mentioned puzzle in the ground-state band of $^{20}$Ne, the negative-parity rotational band built upon the $1^-$ state at $Ex$ = 5.79 MeV which is the inversion-doublet partner of the 
 ground-state band was studied by using the THSR wave functions in Ref.~\cite{Zh13}.   It was found that, also in this 
 negative-parity rotational band, the $^{16}$O + $\alpha$ Brink-GCM wave functions are almost 100$\%$ equivalent to single $^{16}$O + $\alpha$ THSR wave functions.   Thus both rotational bands constituting the inversion doublet are found to have nonlocalized cluster structure of $^{16}$O and $\alpha$. 
 This is an astonishing finding because, as we mentioned above, the existence of the inversion-doublet bands  has been regarded as a clear manifestation of the existence of the localized clustering.  We have to answer the question, 
 ``Is it possible to explain the existence of inversion-doublet bands from the nonlocalized cluster 
 dynamics?''  
 
 The purpose of this paper is to answer the question how the nonlocalized cluster dynamics described by the 
 THSR wave function can be compatible with the concept of a localized cluster structure which has explained the 
 existence of rotational spectra and inversion-doublet bands.  Our purpose is then to propose a new understanding of nuclear 
 cluster dynamics.  We will discuss that the cluster dynamics is of nonlocalized nature but the inter-cluster 
 Pauli repulsion gives rise to the molecular structure of clusters, which in $^{20}$Ne is the $^{16}$O + $\alpha$ 
 molecular structure that generates the inversion-doublet rotational bands.  In order to achieve this goal, 
 we have to solve some problems which we have encountered in the two papers on $^{20}$Ne, Refs.~\cite{Zh12},  \cite{Zh13}, and also in the previous papers on structure studies with the use of THSR wave functions. For example in Refs.~\cite{Zh12,Zh13} it is reported that the THSR wave functions of the ground-state band at the minimum-energy points are of prolate shape while those of the negative-parity band are oblate.  As is mentioned in Ref.~\cite{Zh13}, the oblate THSR wave functions of the negative-parity band are almost 100$\%$ equivalent to respective prolate THSR wave functions.  
This fact is necessary in order to maintain the idea 
that both positive-parity and negative-parity bands are generated from the same intrinsic state. 
 The fact, that after the angular-momentum projection a prolate THSR wave function is almost equivalent to a certain oblate THSR wave function and vice versa, was found and discussed already in the study of $^8$Be~\cite{Fu02} and also in the study of $^{12}$C~\cite{Fu05}.   In this paper we will clarify the reason and the physical meaning of this fact which is general in the case of two-cluster systems.  We will explain that in two-cluster systems 
there exists no physically oblate deformation by showing that after angular-momentum projection even oblate THSR wave functions gives negative quadrupole moments implying that the intrinsic quadrupole moment is of positive sign and hence the intrinsic deformation is prolate.  We will also show that any oblate THSR wave 
 function is equivalent almost 100$\%$ to a rotation average of a prolate THSR wave function around the axis 
 perpendicular to the symmetry axis of prolate deformation. We will calculate the density distribution 
 of prolate THSR wave functions in order to demonstrate undoubtedly that the inter-cluster Pauli repulsion gives rise 
 to molecular structures.  We will see that the THSR wave function which expresses nonlocalized clusters 
 corresponds to the density distribution of the molecular configuration of clusters.   Our new understanding of nuclear cluster 
 dynamics can be stated in saying that nuclear dynamics are basically of nonlocalized nature but the Pauli repulsion makes the two-cluster system look like having effectively localized clustering. Based on the fact that all the cluster 
 states we have ever studied by using THSR wave functions are well described by single THSR wave functions, we know that the clusters make independent motion occupying the lowest orbit of the harmonic-oscillator-like 
 mean-field potential characterized by the size parameter $B$ with a magnitude 
similar to the nuclear radius. We already know that the excitation mode of the system is well described by the Hill-Wheeler equation with the parameter $B$ treated as 
 the Hill-Wheeler coordinate in the systems of 3$\alpha$'s~\cite{To01,Fu03,Fu05} and 4$\alpha$'s~\cite{To01,Fu10}.  
 Therefore we see that the excitation of the system is described  firstly by the dynamics of the size parameter $B$ 
 which is adopted as the generator coordinate and secondly by the excitation of the single-particle motion of clusters 
 in their own mean field potential.   We will call our new understanding of nuclear cluster dynamics the container 
 picture of nuclear clustering, by which we aim to stress that the central quantity of cluster dynamics is the 
 size parameter $B$ of the cluster mean-field potential which we call the container. It is clear that in this picture 
 the existence of cluster-gas states is natural and the formation mechanism of cluster-gas states is just due to the 
 wide extension of the container. 

In order to further facilitate the understanding of the picture we want to promote, it may be helpful for the reader to view the cluster motion in the container as free as long as the clusters do not overlap with one another. This picture which is similar to the concept of excluded volume applies mostly to states of low density (${}^{8}$Be and Hoyle state in ${}^{12}$C). However, as we will see in this paper, also strongly deformed states can easily show cluster structures of this type. It should be kept in mind, however, that this ``container picture" is at its limit for two cluster systems which are strongly prolate, i.e. of molecular structure, but becomes more and more adequate for low density states with more than two clusters when the container is either spherical or not so strongly deformed. 

 The organization of this paper is as follows:  In Sec.~\ref{sec2}, we discuss the hybrid-Brink-THSR wave function and 
 the energy surfaces corresponding to  this wave function.  As was shown in Ref.~\cite{Zh13}, energies are lowest for vanishing 
 inter-cluster distance parameter. In Sec.~\ref{sec3}, we discuss the properties of the wave function with respect to the energy surface, where we discuss large overlaps between prolate and oblate THSR wave functions after angular-momentum projection. In Sec.~\ref{sec4}, we point out that in two-cluster systems there exists no physically oblate deformation 
 and then show that the reason is that any oblate THSR wave function is equivalent almost 100$\%$ to a rotation average of a prolate THSR wave function. In Sec.~\ref{sec5}, we discuss the container picture of nuclear clustering. In that section, we show that the THSR wave functions of two-cluster systems have the density distribution of molecular configuration of clusters, which is caused by the inter-cluster Pauli repulsion. In Sec.~\ref{sec6}, we give our summary with discussions. 

\section{Hybrid-Brink-THSR wave function, preference of zero inter-cluster distance, and the equivalence of the Brink-GCM wave function to a single THSR wave function \label{sec2}}

Let us begin with the original, deformed THSR wave function \cite{Fu02}, which was introduced to describe gas-like $n \alpha$ cluster states with dilute density, 
\begin{equation}
\Phi_{n\alpha}(\vect{\beta})=\int d^3R_1\ldots d^3R_n \exp \Big( -\sum_{i=1}^{n} \sum_{k=x,y,z}
 \frac{R_{ik}^2}{\beta_{k}^2} \Big ) \Phi_{n \alpha}^B(\vect{R}_1,\ldots, \vect{R}_n)
\label{thsr1}
\end{equation}
\begin{equation}
\propto {\cal A}[\prod^{n}_{i=1}\exp \Big(-2\sum_{k=x,y,z} \frac{X_{ik}^2}{B_{k}^2} \Big) \phi(\alpha_{i}) ],
\label{thsr2}
\end{equation}
where 
\begin{eqnarray}
\label{brinkslater}
\Phi_{n \alpha}^B(\vect{R}_1,\ldots, \vect{R}_n)&&=\frac{1}{ \sqrt{ (4n)! }} \text{det}[\phi_{0s}(\vect{r}_1-\vect{R}_1)\chi_{\tau_1,\sigma_1} \cdots \phi_{0s}(\vect{r}_{4n}-\vect{R}_n)\chi_{\tau_{4n},\sigma_{4n}} ] \\
&&\propto  {\cal A}[\prod^{n}_{i=1}\exp\{-2 \frac{( \vect{X}_{i}-\vect{R}_{i})^2}{b^2} \}\phi(\alpha_{i}) ].
\label{brink1}
\end{eqnarray}
Here $B^2_{k}=b^2+2 \beta^2_{k}$, $(k=x,y,z)$, $b$ is the size parameter of the harmonic-oscillator wave function,   $\vect{X}_i$ is the center-of-mass of the $i$-th $\alpha$-cluster $\alpha_i$, and $\phi(\alpha_{i})$ represents the internal wave function of $\alpha_i$.
$\Phi_{n \alpha}^B(\vect{R}_1,\ldots, \vect{R}_n)$ is the $n\alpha$ Brink wave function, which is written as a Slater determinant in Eq.~(\ref{brinkslater}).  $\chi_{\tau,\sigma}$ is the spin-isospin wave function of a nucleon. $\phi_{0s}(\vect{r}-\vect{R})$ is a $0s$ harmonic-oscillator wave function around a center $\vect{R}$  as follows, 
\begin{equation}
\phi_{0s}(\vect{r}-\vect{R})=(\frac{1}{\pi b^2})^{\frac{3}{4}}\exp[-\frac{(\vect{r}-\vect{R})^2}{2b^2}].
\end{equation}
It is to be noted that a new parameter $\vect{\beta}$ (or $\vect{B}$), representing the size of the nucleus, is introduced in the THSR wave function, which is completely different from the parameter $\vect{R}_i$, representing the position of the $i$-th $\alpha$ cluster in the Brink wave function Eq.~(\ref{brink1}). In the THSR wave function, the $n \alpha$ clusters occupy an identical orbit $\exp(-2 X_{x}^2/B_{x}^2-2 X_{y}^2/B_{y}^2-2 X_{z}^2/B_{z}^2)$, and as far as $B_k \gg b$ for $(k=x,y,z)$, the antisymmetrizer ${\cal A}$ is negligible. Then the $n\alpha$ clusters make an independent nonlocalized motion like a gas, in the whole nucleus whose size is characterized by the parameter $\vect{B}$ \cite{Fu09}.

On the other hand, recently, to extend and further clarify the concept of nonlocalized clustering even in non-gas-like cluster states with more compact density, we proposed a new type of microscopic cluster wave function \cite{Zh13}, which we call hybrid-Brink-THSR wave function,   
\begin{equation}
\Phi_{\rm cluster}(\vect{\beta}_i, \vect{S}_i) =\int d^3R_1\ldots d^3R_n 
 \exp\{-\sum_{i=1}^{n} \sum_{k=x,y,z}
\frac{R_{ik}^2}{\beta_{ik}^2} \}   \Phi_{\rm cluster} ^B(\vect{R}_1+\vect{S}_1,\ldots, \vect{R}_n+\vect{S}_n)
\label{hybrid1}
\end{equation}
\begin{equation}
\propto {\cal A}[\prod^{n}_{i=1}\exp\{-A_{i} \sum_{k=x,y,z}{\frac{( X_{ik}-S_{ik})^2}{2 B_{ik}^2}} \}\phi(C_{i}) ],
\label{hybrid2}
\end{equation}
\begin{equation}
\Phi_{\rm cluster} ^B(\vect{S}_1,\ldots, \vect{S}_n) = {\cal A}[\prod^{n}_{i=1}\exp\{-A_{i} \frac{( \vect{X}_{i}-\vect{S}_{i})^2}{2 b^2} \}\phi(C_{i}) ].
\label{brink2}
\end{equation}
Here $\vect{\beta}_i\equiv (\beta_{ix},\beta_{iy},\beta_{iz})$, and $\vect{X}_{i}$ and $\phi(C_{i})$ are the center-of-mass coordinate and the internal wave function of the cluster $C_{i}$, respectively. Different clusters $C_{i}$ can have different mass numbers $A_{i}$ and variational parameters $\vect{\beta}_i$. The oscillator parameter of the cluster $C_i$ is called $b$, which also can be adopted so as to have different values for different clusters.  $\Phi_{\rm cluster}^B$ is the corresponding general Brink model wave function~\cite{brinkcluster}.

In Eq.~(\ref{hybrid1}), another generator coordinate $\vect{S}_i$ is introduced to the original THSR wave function 
Eq.~(\ref{thsr1}). It can be seen from Eq.~(\ref{hybrid2}) that this hybrid wave function combines the important characters of the Brink model as in Eq.~(\ref{brink2}) and the THSR wave function in a very simple way. When $\vect{S}_i$=0, Eq.~(\ref{hybrid1}) corresponds to the THSR wave function and $\vect{\beta}_i$ or $\vect{B}_i$ becomes the size parameter. When $\beta_{ik}= 0$ , i.e., $B_{ik}=b~(k=x,y,z)$, this equation is nothing more than the Brink wave function Eq.~(\ref{brink2}) and $\vect{S}_i$ is the position parameter of the cluster $C_i$. 

As we know, the THSR model provides a nonlocalized clustering picture for the cluster structure rather than the localized clustering represented by the Brink model~\cite{Fu09}.
Since these two different kinds of pictures for clustering are both included in the hybrid-Brink-THSR wave function as the aforementioned two limits, this hybrid wave function provides a very nice way for verifying which picture is more adequate for understanding the relative motions of the cluster structures in nuclei.
    
Now, based on the above hybrid-Brink-THSR wave function, the following cluster wave function of \nene\ can be obtained, as it was considered in Ref~\cite{Zh13},
\begin{equation}
\label{ne1}
\Phi _{\text{Ne}}(\vect{\beta},\vect{S}) =\int d^3R \exp\{-(\frac{4R_{x}^2}{5\beta_{x}^2}+\frac{4R_{y}^2}{5\beta_{y}^2} +\frac{4R_{z}^2}{5\beta_{z}^2})\}
 \Phi_{\text{Ne}} ^B(\frac{4}{5}(\vect{R+S}),-\frac{1}{5}(\vect{R+S}))
\end{equation}
\begin{eqnarray}
\label{ne2}
&&\propto \exp(-\frac{10X_{G}^2}{b^2}) {\widehat \Phi}_{\text{Ne}}(\vect{\beta},\vect{S}),\nonumber \\
&& {\widehat \Phi}_{\text{Ne}}(\vect{\beta},\vect{S}) = {\cal A}[\exp(-\sum_{k=x,y,z}\frac{8(r_k-S_k)^2}{5B_{k}^2})\phi(\alpha)\phi({^{16}\text{O}})],
\end{eqnarray}
where $B^2_{k}=b^2+2 \beta^2_{k}$, $(k=x,y,z)$, $\vect{r}=\vect{X}_{1}-\vect{X}_{2}$, $\vect{X}_G=(4\vect{X}_1+16\vect{X}_2)/20$, and ${\widehat \Phi}_{\text{Ne}}(\vect{\beta},\vect{S})$ is the intrinsic wave function where the spurious center-of-mass motion is eliminated from $\Phi _{\text{Ne}}(\vect{\beta},\vect{S})$ in Eq.~(\ref{ne1}). $\vect{X}_{1}$ and $\vect{X}_{2}$ represent the center-of-mass coordinates of the $\alpha$ cluster and the \oo\ cluster, respectively. All calculations are performed with restriction to axially
symmetric deformation, that is, $\beta_x=\beta_y \neq \beta_z$ and $\vect{S}\equiv(0,0,S_z)$. The spin and parity eigenfunctions can be obtained by the angular-momentum projection technique~\cite{Zh12}, as follows:
\begin{equation}
{\widehat \Phi}^{J^\pi}_{\text{Ne}}(\vect{\beta},\vect{S}) = {\widehat P}^J_{M0} {\widehat \Phi}_{\text{Ne}}(\vect{\beta},\vect{S}), \label{ne3}
\end{equation}
where the parity $\pi=(-1)^J$ and ${\widehat P}^J_{M0}$ is the angular-momentum-projection operator.

The nuclear Hamiltonian we use in this work is given as,
\begin{equation}
{\widehat H}=\sum_{i=1}^{20}T_i - T_G + \sum_{i<j}^{20} (V_{ij}^{(N)} + V_{ij}^{(C)}),
\end{equation}
where the center-of-mass kinetic energy $T_G$ is subtracted from the one-body kinetic term $T_i$. As the nuclear interaction $V_{ij}^{(N)}$, we adopt the same effective nuclear force, Volkov No.1 with the Majorana parameter $M=0.59$, and the same oscillator parameter $b=1.46$ fm, as were used in our previous papers~\cite{Zh12,Zh13}. $V_{ij}^{(C)}$ is the Coulomb interaction between protons.

\begin{figure}[!h]
\centering
\includegraphics[scale=0.42]{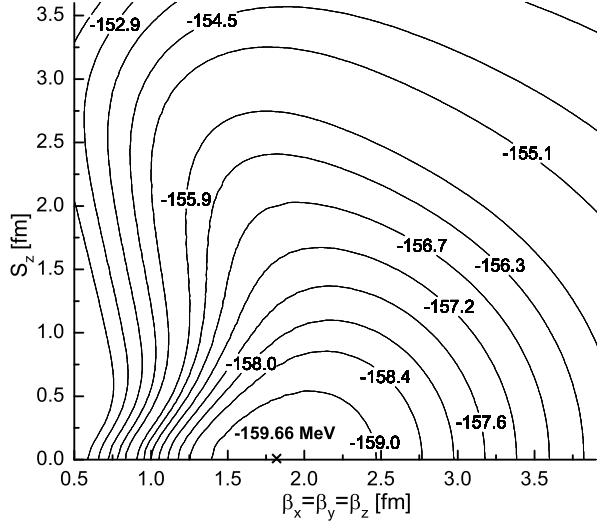}
\caption{\label{intrinsic_0+_eng} Contour map of the energy surface of the intrinsic wave function of \nene\ in the two-parameter space, $S_z$ and $\beta_{x}=\beta_{y}=\beta_{z}$.}
\end{figure}
\begin{figure}[!h]
\centering
\includegraphics[scale=0.42]{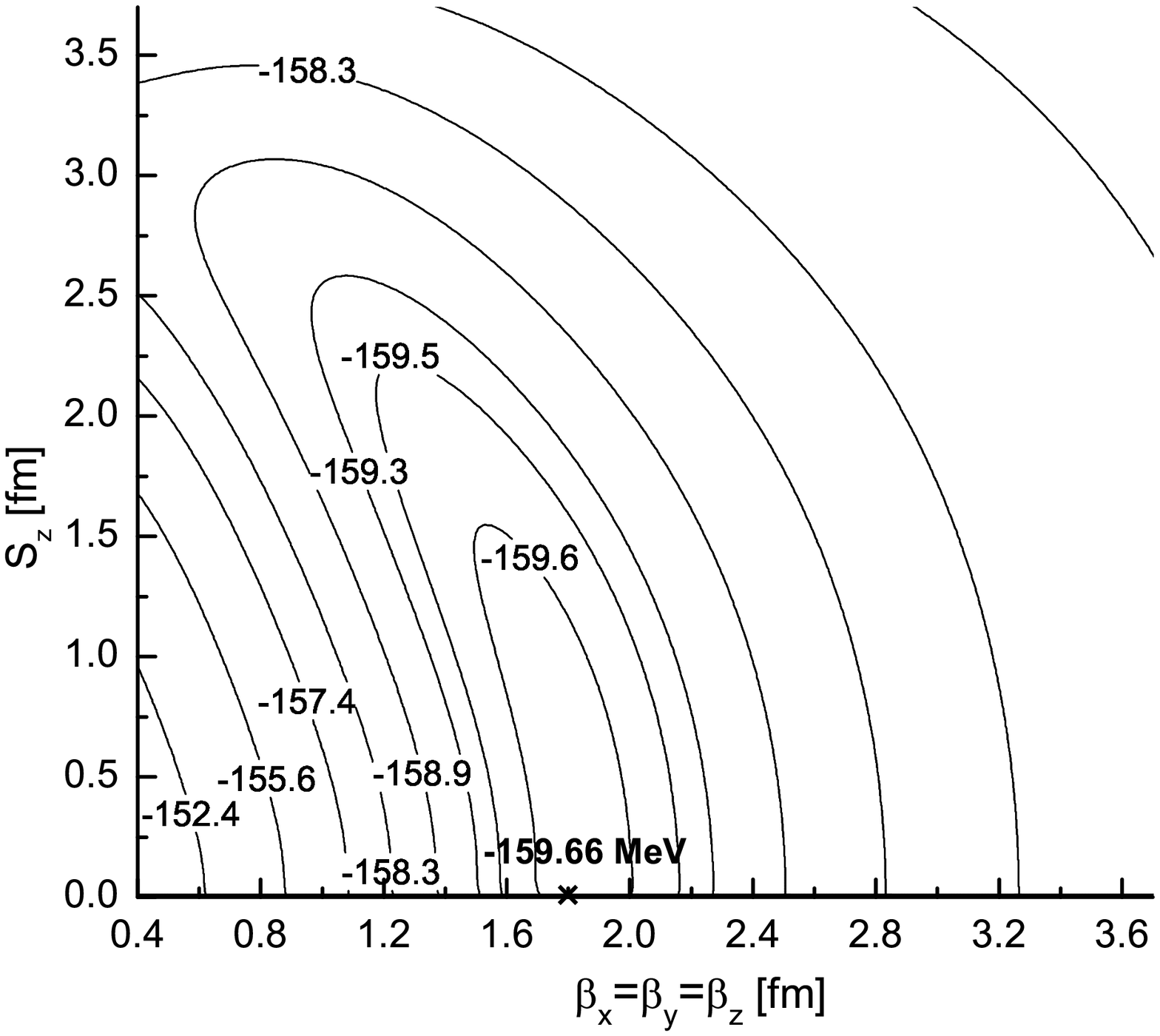}
\caption{\label{0+_eng} Contour map of the energy surface of the  $J^\pi=0^+ $ state in the two-parameter space, $S_z$ and $\beta_{x}=\beta_{y}=\beta_{z}$.}
\end{figure}

\begin{figure}[!h] 
\centering
\includegraphics[scale=0.42]{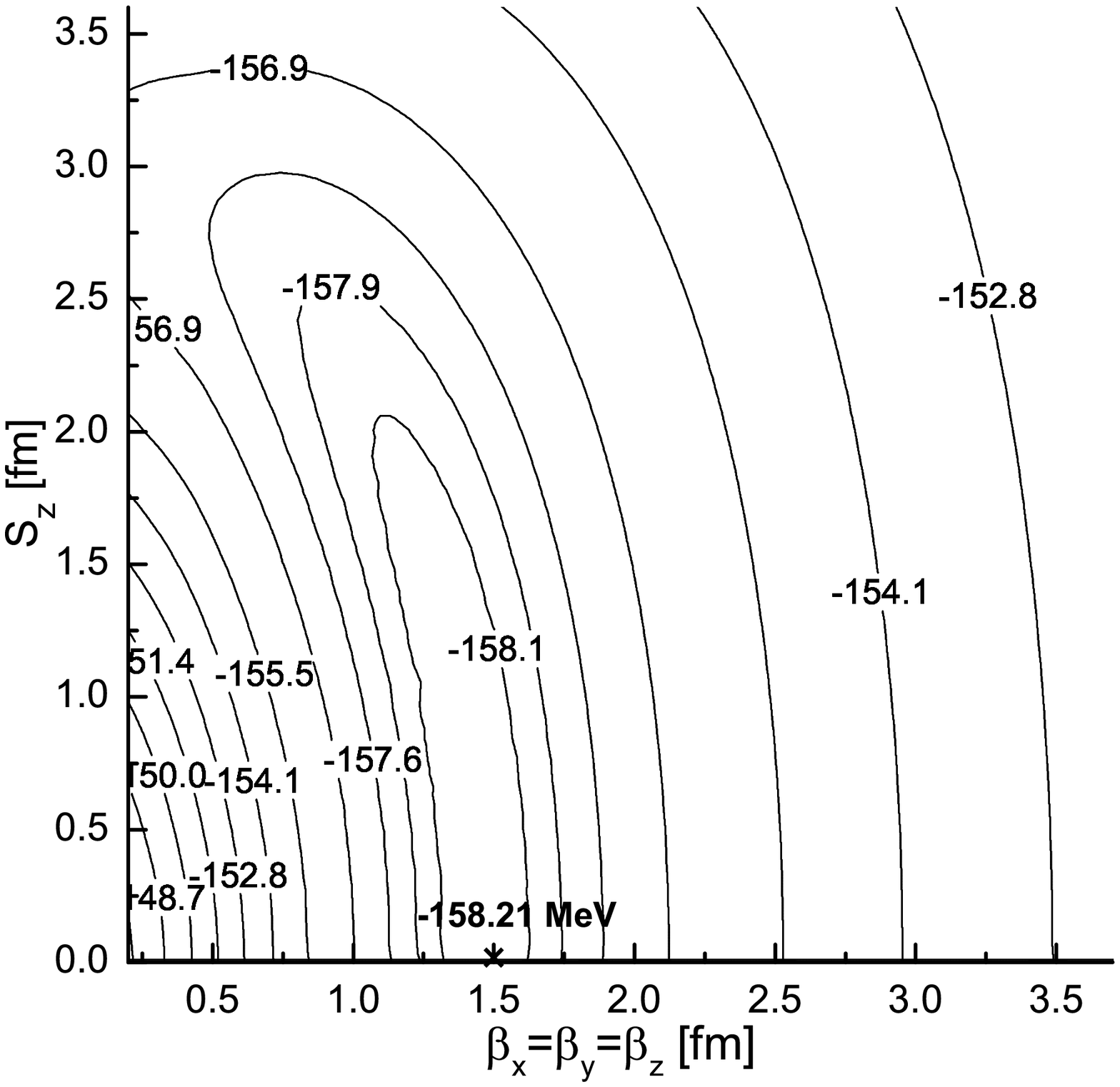}
\caption{\label{2+_eng} Contour map of the energy surface of the  $J^\pi=2^+ $ state in the two-parameter space, $S_z$ and $\beta_{x}=\beta_{y}=\beta_{z}$.}
\end{figure}

\begin{figure}[!h]
\centering
\includegraphics[scale=0.42]{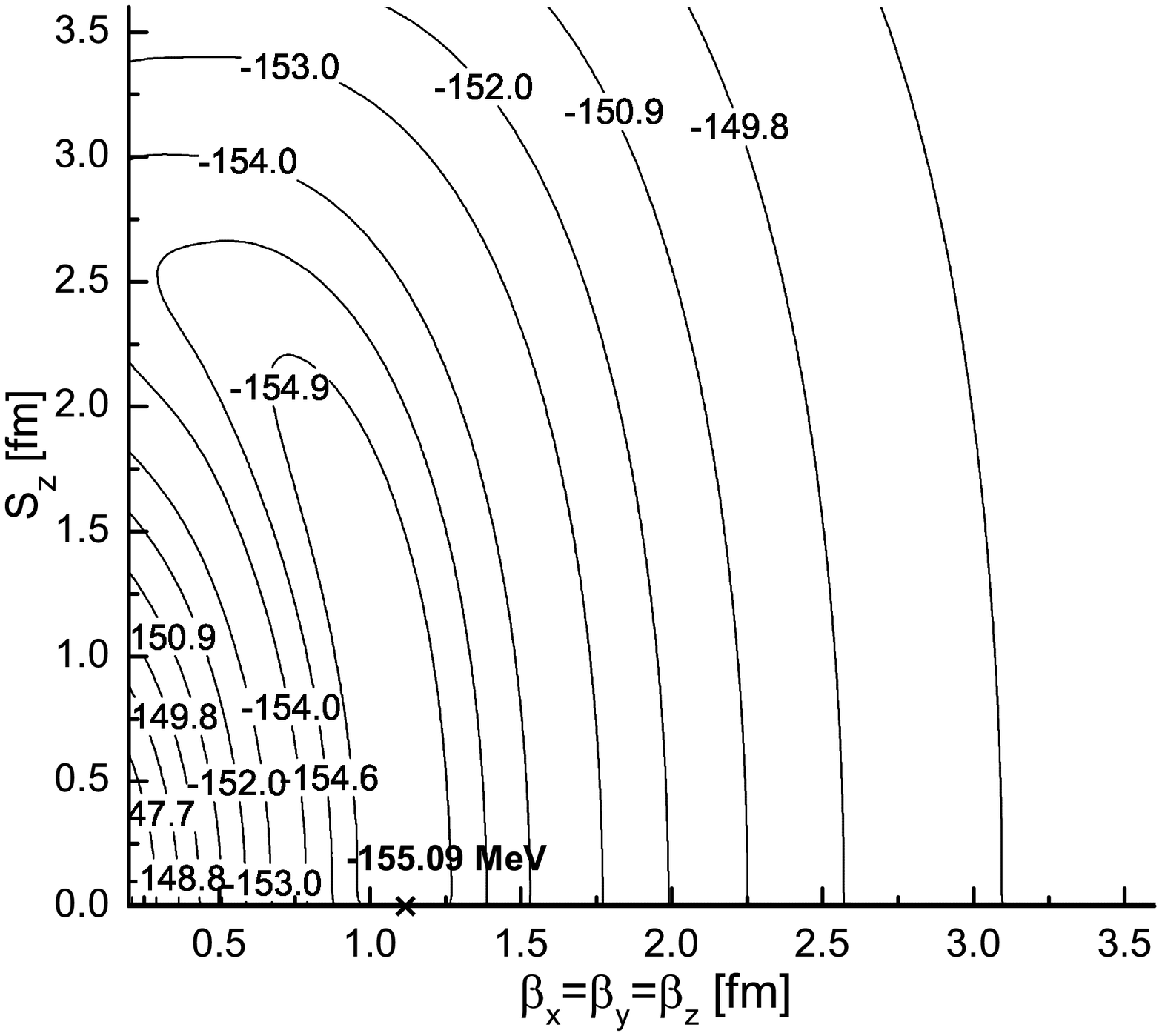}
\caption{\label{4+_eng} Contour map of the energy surface of the  $J^\pi=4^+ $ state in the two-parameter space, $S_z$ and $\beta_{x}=\beta_{y}=\beta_{z}$.}
\end{figure} 

\begin{figure}[!h]
\centering
\includegraphics[scale=0.42]{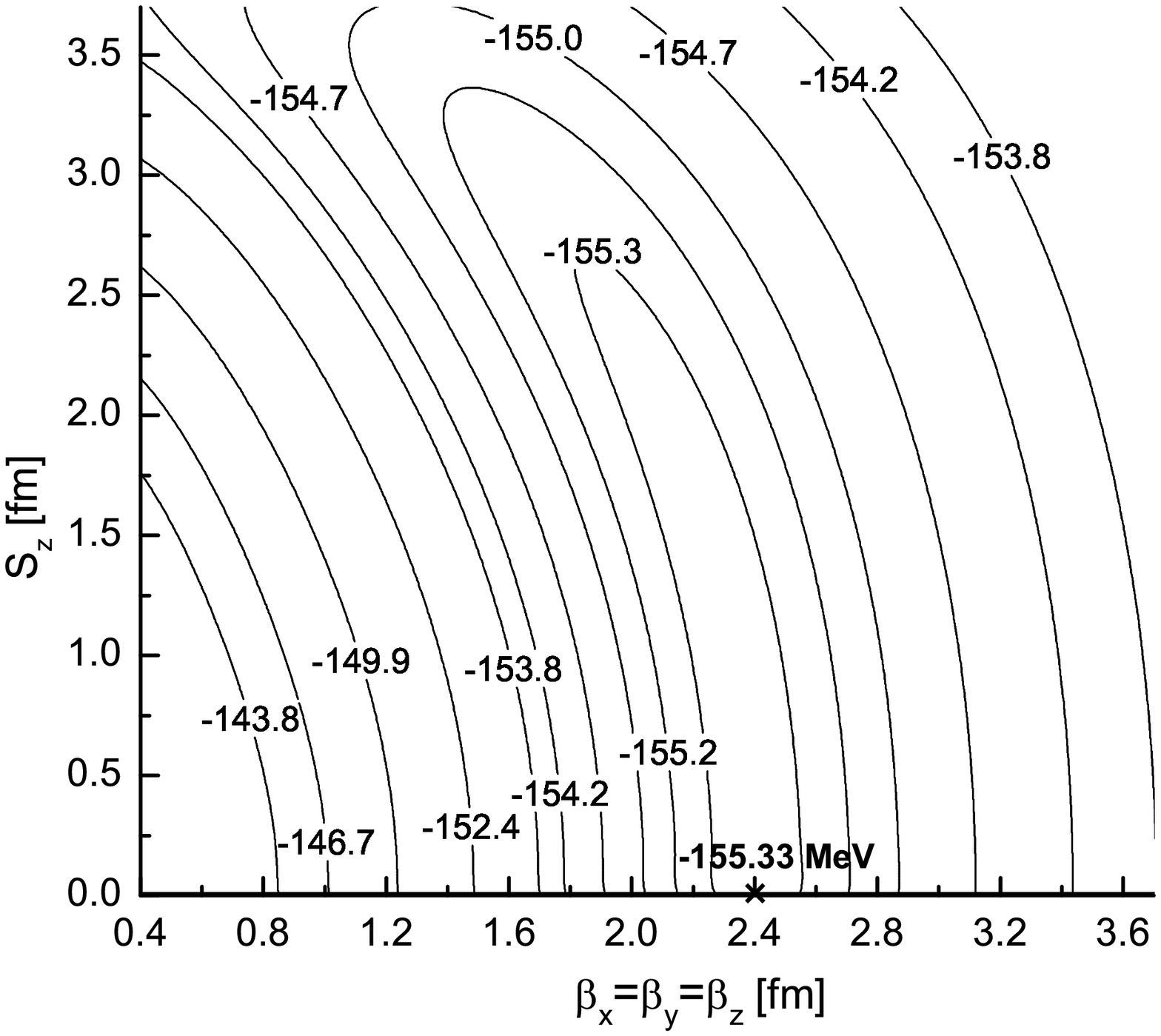}
\caption{\label{1-_eng} Contour map of the energy surface of the  $J^\pi=1^- $ state in the two-parameter space, $S_z$ and $\beta_{x}=\beta_{y}=\beta_{z}$.}
\end{figure}

\begin{figure}[!h]
\centering
\includegraphics[scale=0.42]{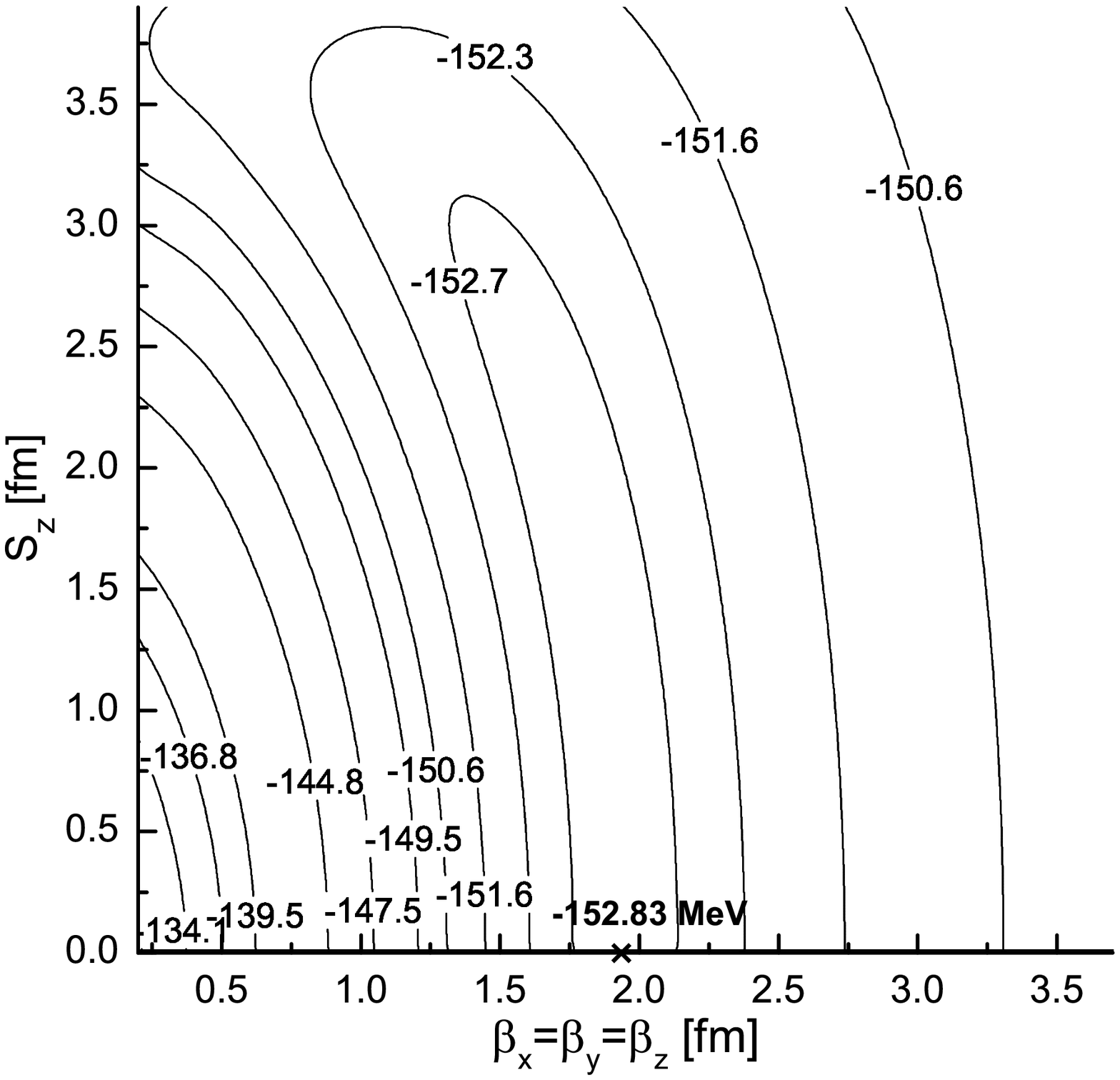}
\caption{\label{3-_eng} Contour map of the energy surface of the  $J^\pi=3^- $ state in the two-parameter space, $S_z$ and $\beta_{x}=\beta_{y}=\beta_{z}$.}
\end{figure}

\begin{figure}[!h]
\centering
\includegraphics[scale=0.42]{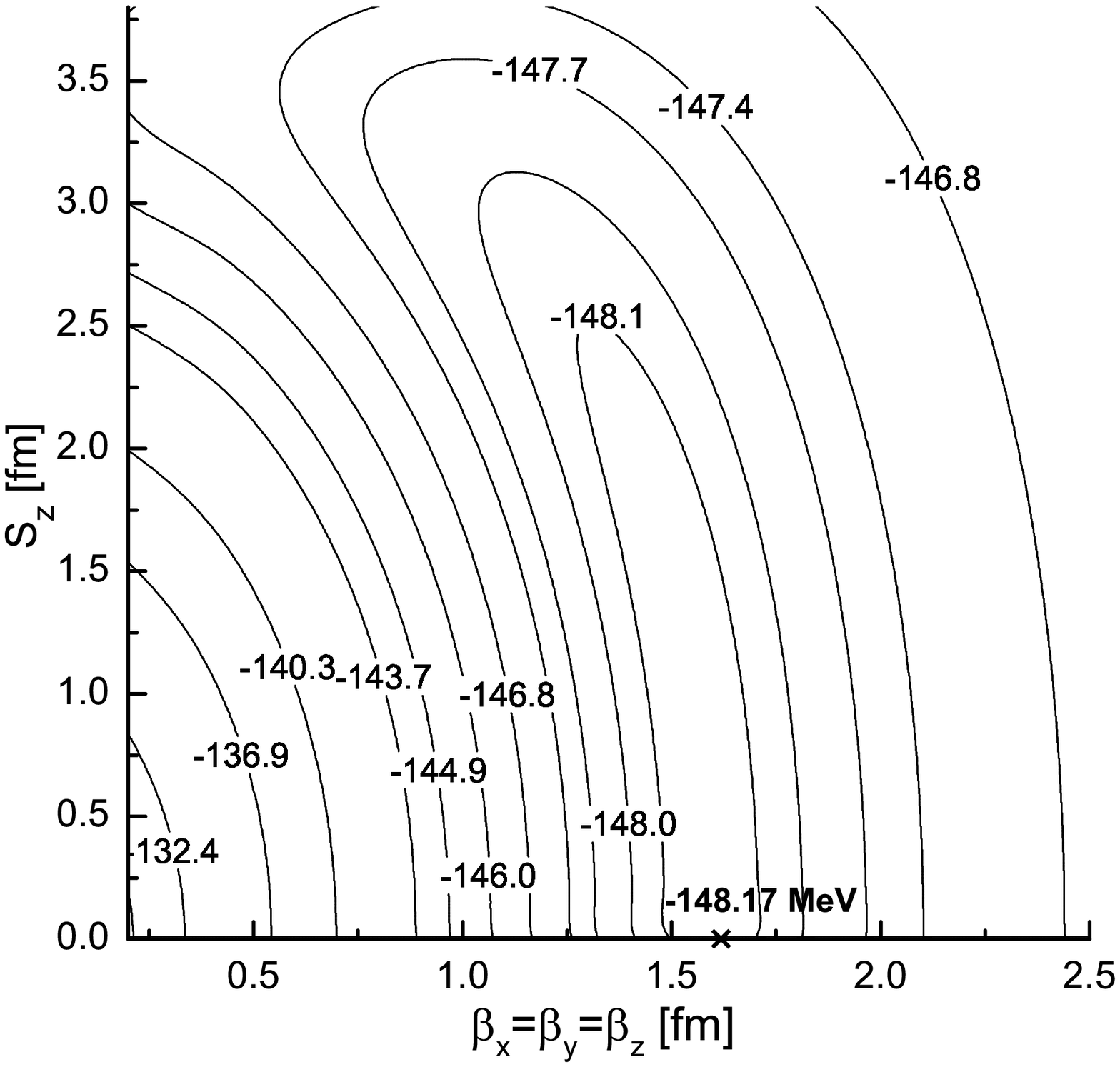}
\caption{\label{5-_eng} Contour map of the energy surface of the  $J^\pi=5^- $ state in the two-parameter space, $S_z$ and $\beta_{x}=\beta_{y}=\beta_{z}$.}
\end{figure} 

Fig.~\ref{intrinsic_0+_eng} shows the contour map of the energy surface of the intrinsic wave function of \nene\ in the two-parameter space, $S_z$ and $\beta_{x}=\beta_{y}=\beta_{z}$, i.e. the following quantity:
\begin{equation}
E(\beta_x=\beta_y=\beta_z,S_z)=\frac{\langle {\widehat \Phi}_{\text{Ne}}(\beta_x=\beta_y=\beta_z,S_z)|{\widehat H}|
{\widehat \Phi}_{\text{Ne}}(\beta_x=\beta_y=\beta_z,S_z) \rangle}{\langle {\widehat \Phi}_{\text{Ne}}(\beta_x=\beta_y=\beta_z,S_z) | {\widehat \Phi}_{\text{Ne}}(\beta_x=\beta_y=\beta_z,S_z) \rangle}.
\end{equation}
We see that the minimum energy $-159.66$ MeV appears at $S_z$=0 and $\beta_{x}=\beta_{y}=\beta_{z}=1.8$ fm. The value $S_z$=0  means the intrinsic hybrid-Brink-THSR wave function becomes the intrinsic THSR wave function in describing the ground state of \nene. This result indicates that the intrinsic THSR wave function based on the nonlocalized clustering is more suitable for describing the ground state of \nene\ than the Brink wave function.

Next, we perform angular-momentum projection following Eq.~(\ref{ne3}) on the intrinsic hybrid-Brink-THSR wave function ${\widehat \Phi}_{\text{Ne}}(\vect{\beta},\vect{S})$ in Eq.~(\ref{ne2}) for the spherical case $\beta_x=\beta_y=\beta_z$, and then make the following variational calculations for obtaining the optimum wave functions,
\begin{equation}
E^{J^\pi}(\beta_x=\beta_y=\beta_z,S_z)=\frac{\langle {\widehat \Phi}^{J^\pi}_{\text{Ne}}(\beta_x=\beta_y=\beta_z,S_z)|{\widehat H}|
{\widehat \Phi}^{J^\pi}_{\text{Ne}}(\beta_x=\beta_y=\beta_z,S_z) \rangle}{\langle {\widehat \Phi}^{J^\pi}_{\text{Ne}}(\beta_x=\beta_y=\beta_z,S_z) | {\widehat \Phi}^{J^\pi}_{\text{Ne}}(\beta_x=\beta_y=\beta_z,S_z)\rangle}.
\end{equation}
 Figs.~\ref{0+_eng}--\ref{5-_eng} show the contour maps of the above quantity for the different $J^\pi$ states of the inversion doublet bands in \nene\, in the two-parameter space, $S_z$ and $\beta_{x}=\beta_{y}=\beta_{z}$.
It is surprising to find that the obtained minimum energies with respect to the projected states all appear at $S_z$=0. For instance, the obtained minimum energy $-159.66$ MeV for $J^\pi=0^+$ state appears at $S_z=0$ and  $\beta_{x}=\beta_{y}=\beta_{z}=1.8$ fm. For the $J^\pi=1^-$ state,  the minimum energy $-155.33$ MeV appears at $S_z=0$ and  $\beta_{x}=\beta_{y}=\beta_{z}=2.4$ fm in the contour map.
The inter-cluster distance parameter $S_z=0$ means that this hybrid-Brink-THSR wave function tends to a pure THSR wave function in describing the cluster states of the inversion doublet bands in \nene.

It should be noted that although $S_z$ does not give any contribution to the energy gain, it still plays an important role in providing negative-parity states. This is because the parameter $S_z$ is the only component which breaks the parity symmetry, as is clearly seen in the form of the intrinsic wave function ${\widehat \Phi}_{\text{Ne}}(\vect{\beta},\vect{S})$ in Eq.~(\ref{ne2}). In the following, we demonstrate that the negative-parity states can be constructed even in the limiting situation, $S_z\rightarrow0$, for the $\beta_x=\beta_y=\beta_z$ case, for simplicity.
The angular-momentum projected hybrid-Brink-THSR wave function can then be written as,
\begin{equation}
{\widehat P}_{M0}^L {\widehat \Phi}_{\text{Ne}}(\beta_x=\beta_y=\beta_z,S_z) 
    \propto  {\cal A}\Big[j_L(2i \gamma S_z r)Y_{LM}(\widehat{r})e^{-\gamma r^2}\phi(\alpha)\phi(^{16}\text{O})\Big]
\label{angpro1}
\end{equation}
\begin{equation}
 \propto S_z^L \Phi_{LM}^{(0)} + {\cal O}(S_z^{L+2}),  
\label{angpro2}
\end{equation}
\begin{equation}
 \Phi_{LM}^{(0)} = {\cal A}\Big[ r^L e^{-\gamma r^2} Y_{LM}(\widehat{r})\phi(\alpha) \phi(^{16}\text{O})\Big].\label{angpro3}
\end{equation}
Here $\gamma=8/(5B^2)$ with $B^2=B_x^2=B_y^2=B_z^2$. Now the normalized and projected wave function of Eq.~(\ref{angpro3}) can be obtained analytically, in the limit of $S_z \rightarrow 0$, as follows: 
\begin{equation}
\frac{\Phi_{LM}^{(0)}}{\sqrt{\langle \Phi_{LM}^{(0)} | \Phi_{LM}^{(0)} \rangle}} = \lim_{S_z \rightarrow 0} \frac{{\widehat P}_{M0}^L {\widehat \Phi}_{\text{Ne}}(\beta_x=\beta_y=\beta_z,S_z)}{\sqrt{\langle {\widehat P}_{M0}^L {\widehat \Phi}_{\text{Ne}}(\beta_x=\beta_y=\beta_z,S_z)|{\widehat P}_{M0}^L {\widehat \Phi}_{\text{Ne}}(\beta_x=\beta_y=\beta_z,S_z)\rangle}}.
\label{anlyticalsz0}
\end{equation}

\begin{figure}[!h]
\centering
\includegraphics[scale=0.42]{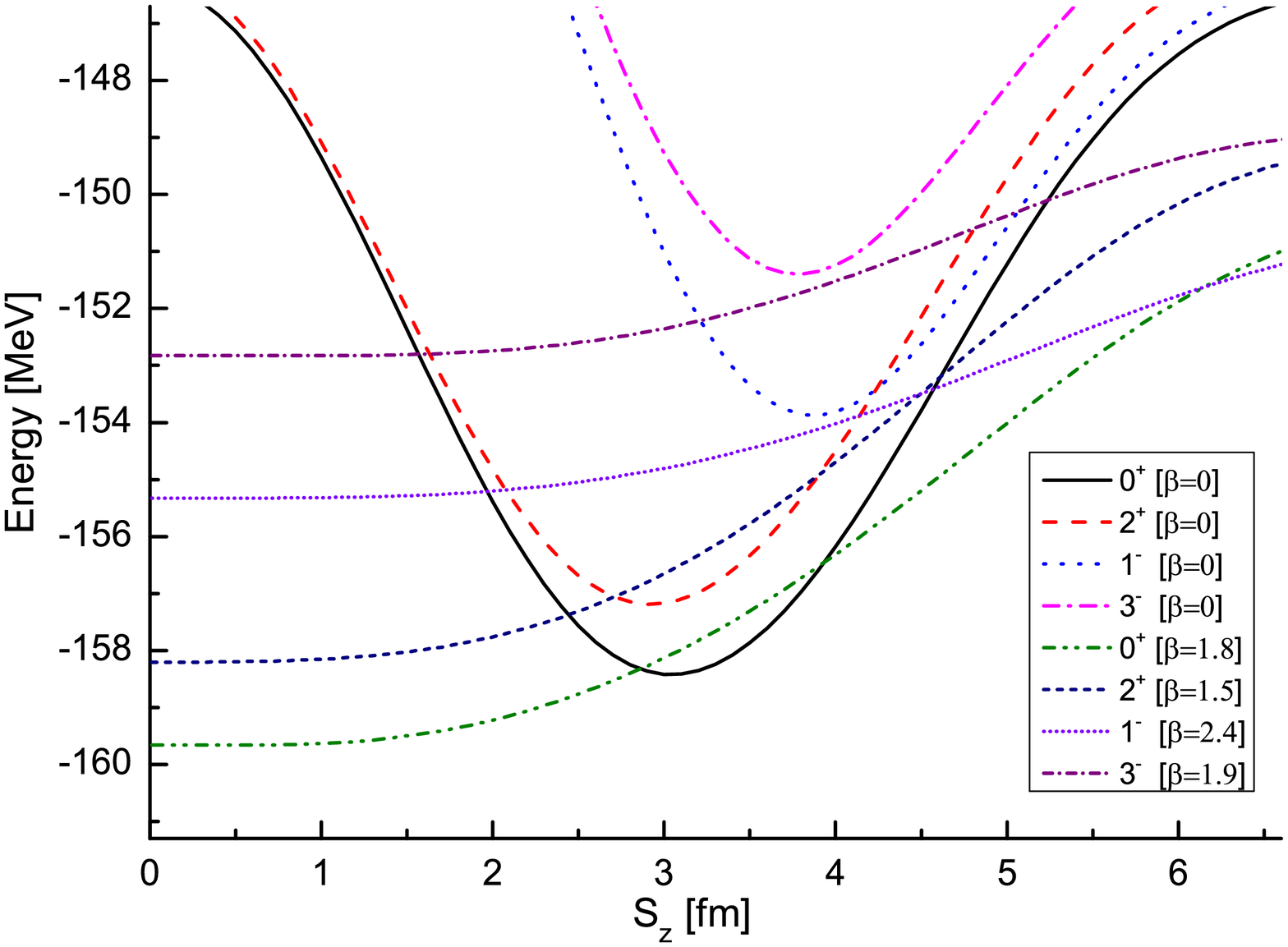}
\caption{\label{nonlocalized}  Energy curves of $J^\pi=0^+$,  $2^+$,  $1^-$,  and $3^-$ states with different widths of Gaussian relative wave functions in the hybrid model.}
\end{figure}  

The above variational calculations with the hybrid-Brink-THSR wave function put the parameter $S_z$ to zero   for the inversion doublet bands. This means that, in spite of the fact that a pure Brink wave function gives  a distinct energy minimum point with non-zero $S_z$, the localized clustering picture cannot be supported. We can realize this situation by looking at Fig.~\ref{nonlocalized}, which shows the energy curves of the lower excited states of the inversion doublet bands with different widths of the Gaussian relative wave functions in the hybrid model. 
If $\vect{\beta}$ is fixed at 0, the hybrid-Brink-THSR wave function becomes the Brink wave function. In this case, $S_z$ is the inter-cluster distance parameter and it is usually regarded as a dynamics parameter for describing the cluster system. For instance, the minimum energy of the ground state of \nene\ appears at $S_z=3.0$ fm. For the $J^\pi=1^-$ state, the optimum position appears at $S_z=3.9$ fm. The non-zero values of $S_z$ seem to indicate that the $\alpha$+\oo\ structure of \nene\  favours localized clustering. This is just the traditional concept of localized clustering. Now, we believe that this argument is misleading \cite{Zh13}.  
The non-zero minimum point $S_z$ simply occurs since the width of the Gaussian wave function of the relative motion in the Brink model is fixed to a narrow wave packet, characterized by the parameter $b$. If we take non-zero values for $\vect{\beta}$, namely, $\beta_x=\beta_y=\beta_z $=$1.8$ fm, $1.5$ fm, $2.4$ fm, and $1.9$ fm for $J^\pi=0^+$, $2^+$, $1^-$, and $3^-$ states, respectively, according to their minimum positions in the contour maps, then we find that the minimum points appear at $S_z=0$ in Fig.~\ref{nonlocalized}.
This indicates that the separation distance parameter $S_z$ does not play any physical role in describing the $\alpha$+\oo\ cluster structure, even for the negative-parity states. Instead of that, the new parametrization by $\vect{\beta}$, which characterizes nonlocalized clustering, is more appropriate for describing the cluster structure in \nene.  

\begin{table}[htbp]
\centering
\caption{\label{table1}$E_{\text{min}}^{J^\pi}(\beta_x=\beta_y, \beta_z)$ are the minimum energies at the corresponding values of $\beta_x=\beta_y$ and $\beta_z$ in the hybrid model.  The squared overlaps between the single normalized projected THSR-type wave functions $ \hat{\Phi}_{\text{Min}}^{\text{THSR}}$ corresponding to the minimum energies and the normalized Brink  GCM wave functions are also listed. Units of energies are MeV. }
\begin{tabular}{ c c c c c c c c}
\hline
\hline
State   & $E_{\text{min}}^{J^\pi}(\beta_x=\beta_y, \beta_z)$& $|\langle\hat{\Phi}_{\text{Min}}^{\text{THSR}}|\hat{\Phi}_{\text{GCM}}^{\text{Brink}} \rangle|^2$  \\ \hline
$0^+$&-159.85(0.9, 2.5)&0.9929 \\
$2^+$&-158.53(0.0, 2.2)&0.9879 \\
$4^+$&-155.50(0.0, 1.8)&0.9775\\
$1^-$&-155.38(3.7, 1.4)&0.9998 \\
$3^-$&-153.07(3.7, 0.0)&0.9987 \\
 \hline
 \hline
 \end{tabular}
\end{table} 
 
Now that the parameter $S_z$ tends to be zero in the obtained hybrid wave functions of the inversion doublet bands in \nene,  we can make further variational calculations in the two-parameter space, $\beta_{x}=\beta_{y}$ and $\beta_{z}$  using the projected hybrid-Brink-THSR wave function  with the parameter $S_z=0$ (In practical calculations $S_z$ is fixed at a very small value close to zero).  We can write the formula as follows,
\begin{equation}
\label{contour}
E^{J^\pi}(\beta_x=\beta_y, \beta_z)=\frac{\langle {\widehat \Phi}^{J^\pi}_{\text{Ne}}(\beta_x=\beta_y, \beta_z)|{\widehat H}|
{\widehat \Phi}^{J^\pi}_{\text{Ne}}(\beta_x=\beta_y, \beta_z)\rangle}{\langle {\widehat \Phi}^{J^\pi}_{\text{Ne}}(\beta_x=\beta_y, \beta_z) | {\widehat \Phi}^{J^\pi}_{\text{Ne}}(\beta_x=\beta_y,\beta_z)\rangle}.
\end{equation}
Here $  {\widehat \Phi}^{J^\pi}_{\text{Ne}}(\beta_x=\beta_y, \beta_z)\equiv {\widehat \Phi}^{J^\pi}_{\text{Ne}}(\beta_x=\beta_y, \beta_z,S_z\rightarrow0). $  The obtained minimum energies and the corresponding values of $\beta_x=\beta_y$ and $\beta_z$ using the THSR-type wave functions are listed  in Table \ref{table1}. 
 
On the other hand, the exact solution of the $\alpha$+\oo\ cluster system can be obtained by superposing the single Brink wave functions,  that is the Brink-GCM wave function.  
\begin{equation}
\label{hillwheeler}
\sum_{j}\langle{\Phi}_{\text{Brink}}^{J^\pi}(R_i)|\widehat{H}-E|{\Phi}_{\text{Brink}}^{J^\pi}(R_j)\rangle f(R_j)=0.
\end{equation}
Here, ${\Phi}_{\text{Brink}}^{J^{\pi}}(R_i)$ can be obtained directly from the projected Brink wave function 
$\Phi_{\text{Brink}}^{J^{\pi}}(\frac{4}{5}\vect{R},-\frac{1}{5}\vect{R})$ with $\vect{R}=(0,0,R_i)$.
Thus, by solving the Hill-Wheeler equation Eq.~(\ref{hillwheeler}),  we can obtain the following Brink-GCM wave function, 
\begin{equation}
\Phi_{\text{GCM}}^{J^{\pi}}=\sum_{i} f(R_i) {\Phi}_{\text{Brink}}^{J^\pi}(R_i).
\end{equation}
Thus, we can compare the single THSR-type wave function with the exact Brink-GCM wave function \cite{Zh12} for the description of the $\alpha$+\oo\ cluster system by calculating their squared overlap $|\langle \hat{\Phi}_{\text{min}}^{\text {THSR}}| \hat{\Phi}_{\text{GCM}}^{\text{Brink}}\rangle|^2$.   
In Table \ref{table1}, we can find that the obtained single THSR-type wave functions have  99.29$\%$, 98.79$\%$,  97.75$\%$, 99.98$\%$,  and 99.87$\%$  squared overlaps for $J^\pi$=$0^+, 2^+$, $4^+$, $1^-$, and $3^- $ states of \nene, respectively, with the corresponding Brink-GCM solutions. These high squared overlaps mean that the single THSR-type wave functions are almost 100$\%$ equivalent to the corresponding RGM/GCM wave functions, thus,  these obtained single  angular-momentum projected THSR-type wave functions can accurately describe the states of the inversion doublet bands in \nene\ \cite{Zh12, Zh13}. Moreover, the concept of nonlocalized clustering proposed by the THSR-type wave function obtained from the hybrid-Brink-THSR wave function is essential to correctly understand the $\alpha+$\oo\ cluster structure in \nene.  In conclusion, we can say that the $S_z$-parameter in the hybrid wave function only serves to sort out even and odd parities. The limiting process $S_z \rightarrow 0$ is very similar to the way with which one obtains from an antisymmetrized product of two Gaussians (S-waves) a P-wave harmonic-oscillator wave function. One first slightly displaces the centers of the Gaussians, then antisymmetrises and normalises and then takes the limit of displacement to zero (see Eq.~\ref{anlyticalsz0}).

\section{Equivalence of prolate and oblate THSR wave functions after angular-momentum projection  \label{sec3}}

\begin{figure}[!h]
\centering
\includegraphics[scale=0.42]{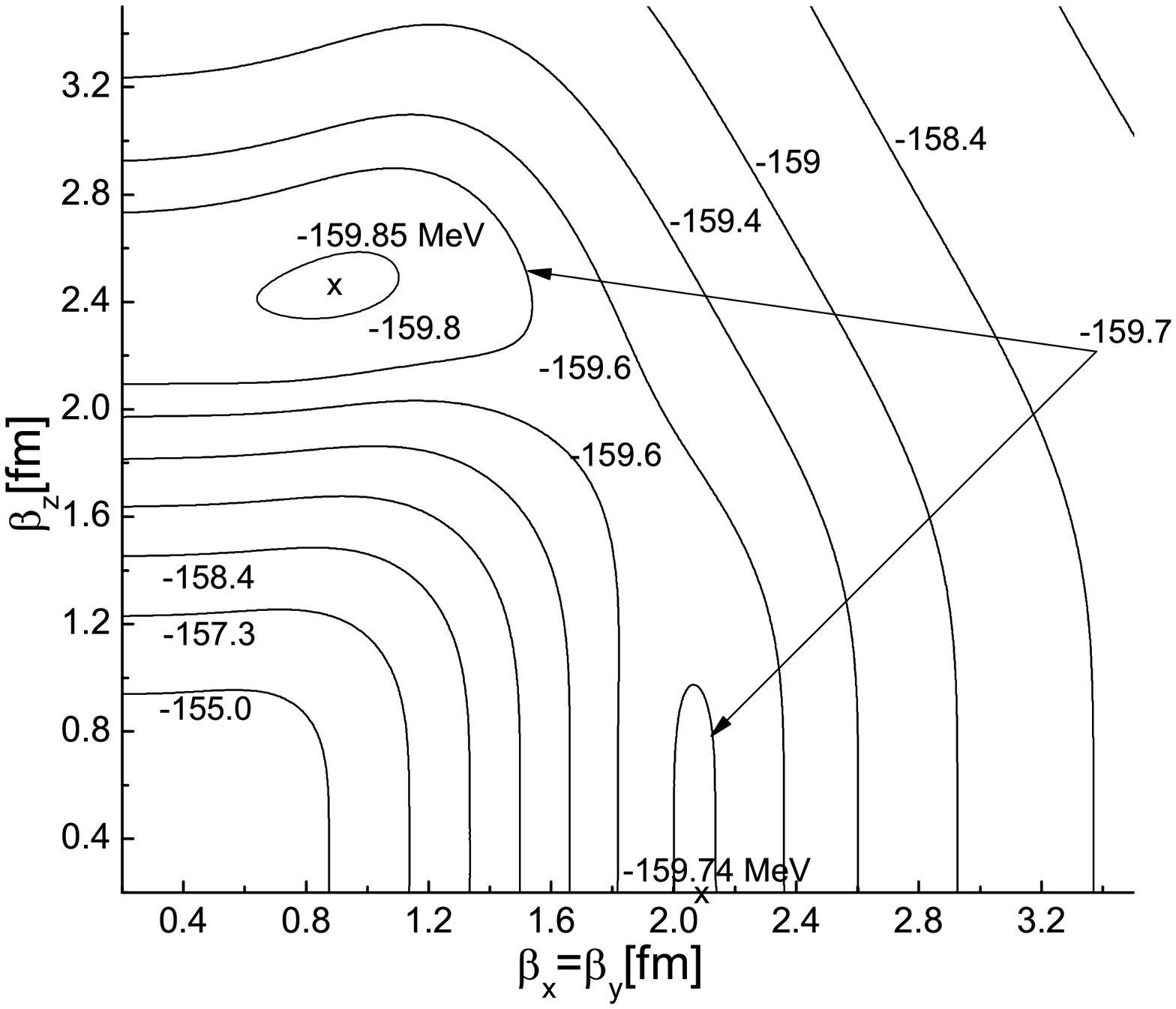}
\caption{\label{0+detail} Contour map of the energy surface of the  $J^\pi=0^+ $ state in the two-parameter space, $\beta_{x}=\beta_{y}$ and $\beta_{z}$.}
\end{figure}

\begin{figure}[!h]
\centering
\includegraphics[scale=0.42]{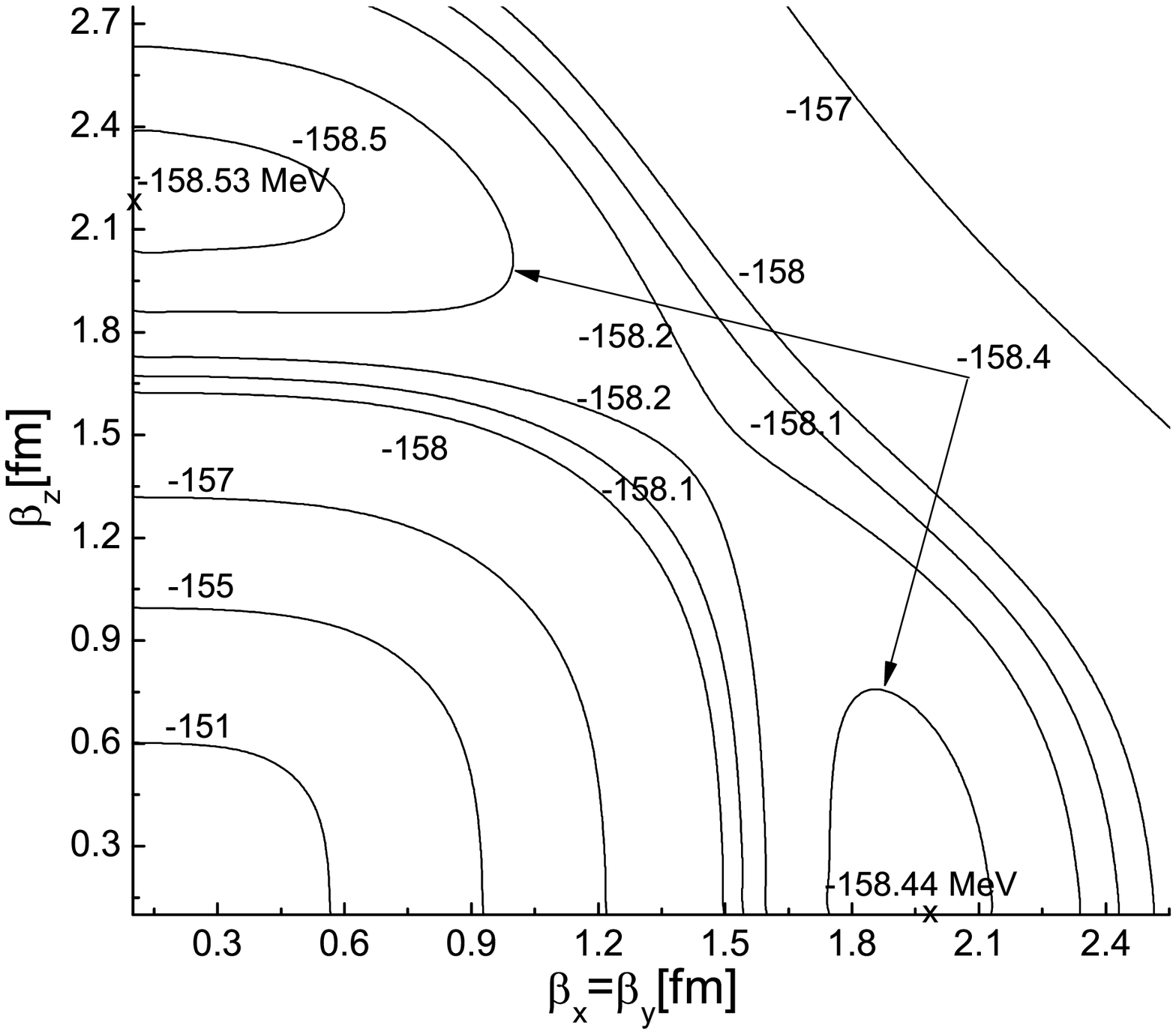}
\caption{\label{2+detail} Contour map of the energy surface of the  $J^\pi=2^+ $ state in the two-parameter space, $\beta_{x}=\beta_{y}$ and $\beta_{z}$.}
\end{figure}

\begin{figure}[!h]
\centering
\includegraphics[scale=0.42]{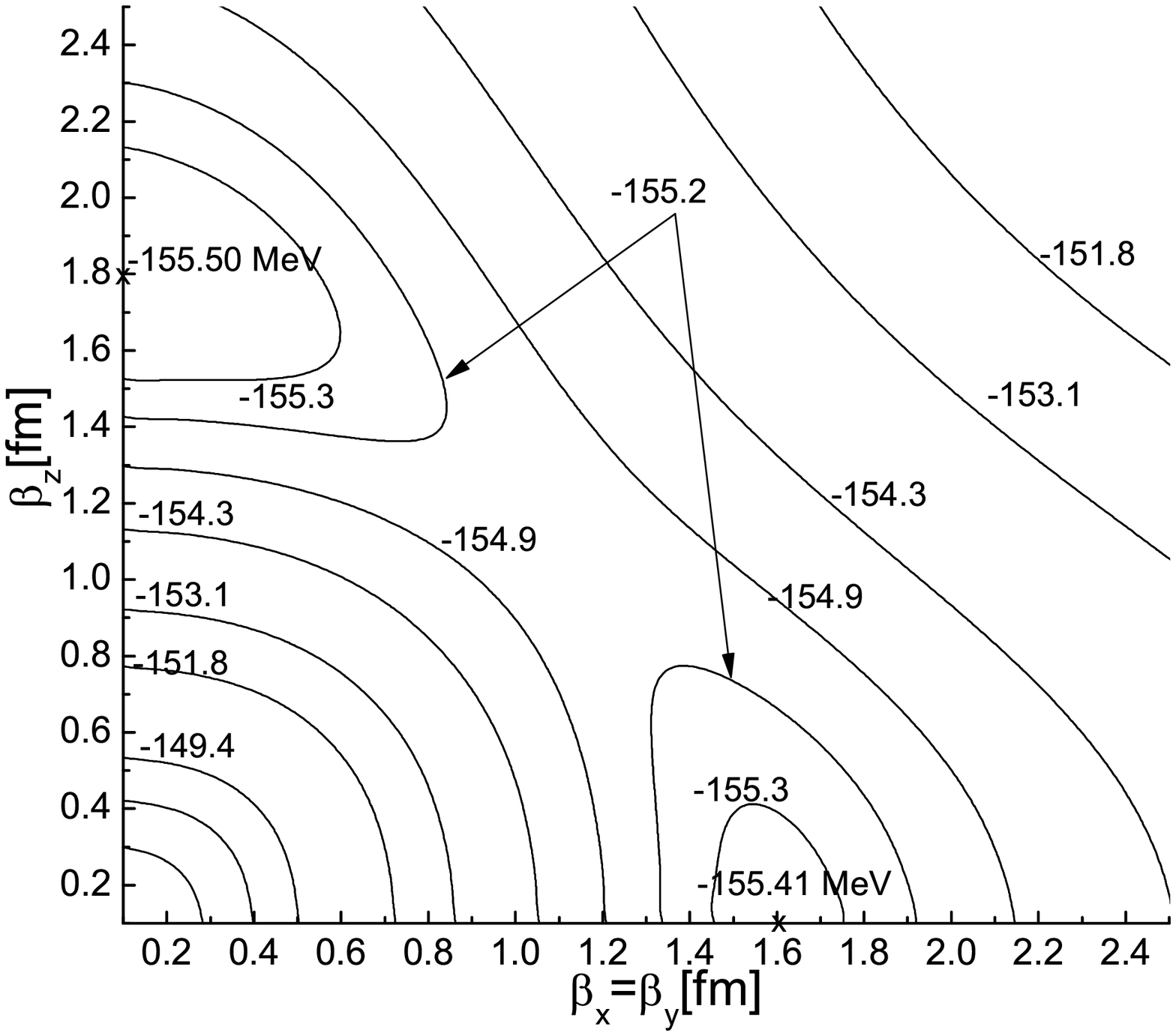}
\caption{\label{4+detail} Contour map of the energy surface of the  $J^\pi=4^+ $ state in the two-parameter space, $\beta_{x}=\beta_{y}$ and $\beta_{z}$.}
\end{figure}

 \begin{figure}[!h]
\centering
\includegraphics[scale=0.42]{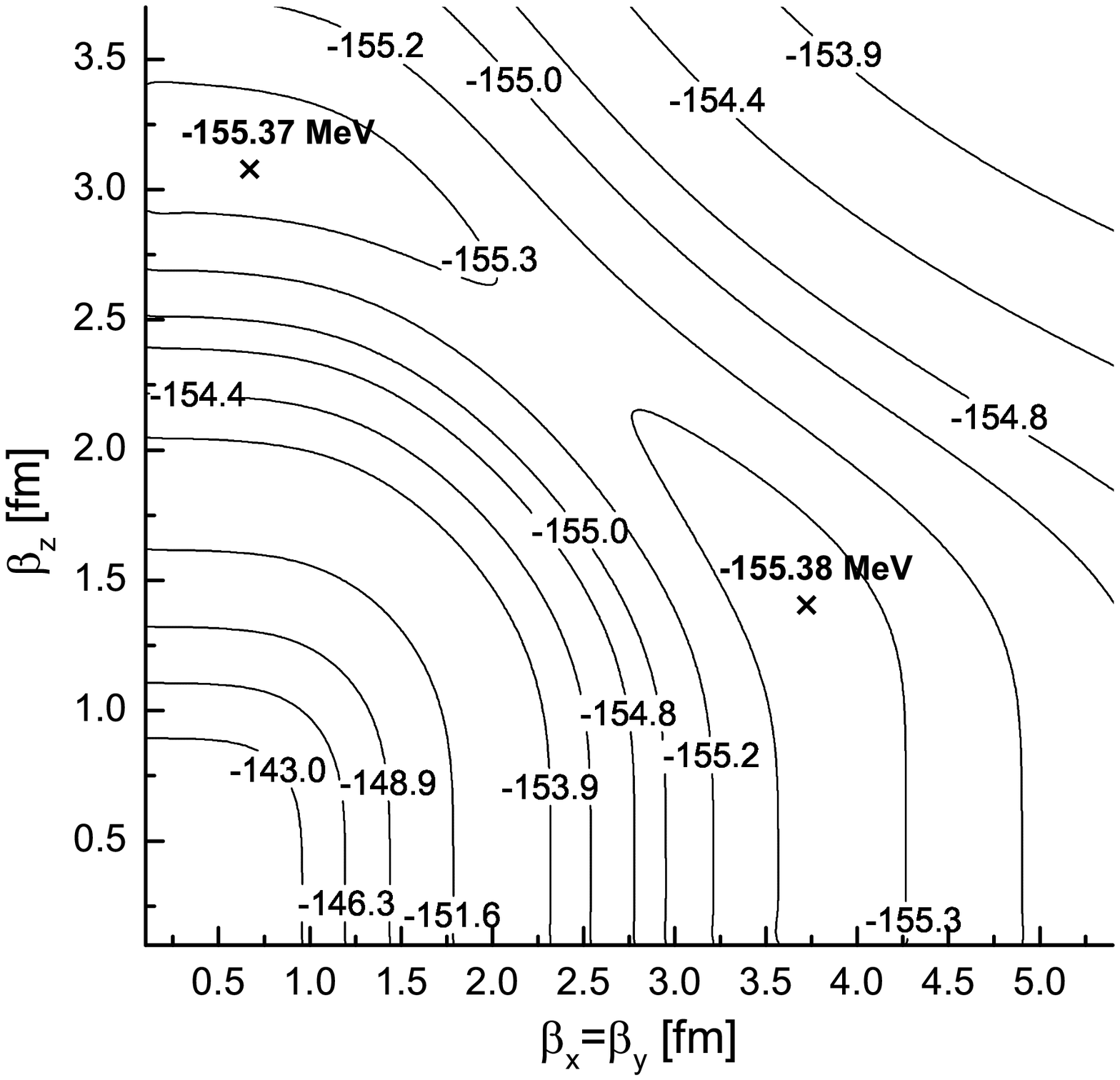}
\caption{\label{1-bxbz} Contour map of the energy surface of the  $J^\pi=1^- $ state in the two-parameter space, $\beta_{x}=\beta_{y}$ and $\beta_{z}$.}
\end{figure}
  \begin{figure}[!h]
\centering
\includegraphics[scale=0.42]{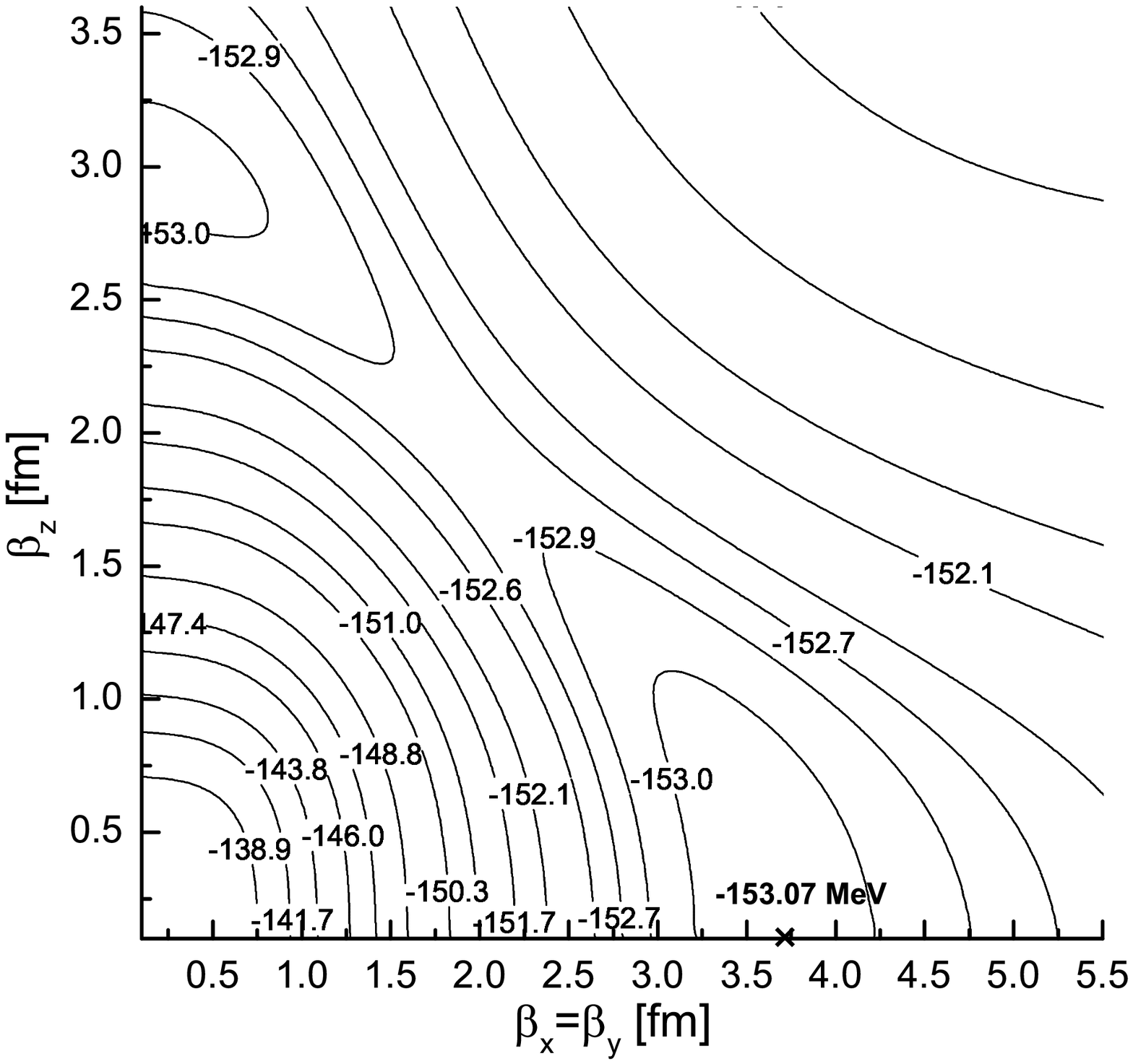}
\caption{\label{3-bxbz} Contour map of the energy surface of the  $J^\pi=3^- $ state in the two-parameter space, $\beta_{x}=\beta_{y}$ and $\beta_{z}$.}
\end{figure}

  \begin{figure}[!h]
\centering
\includegraphics[scale=0.42]{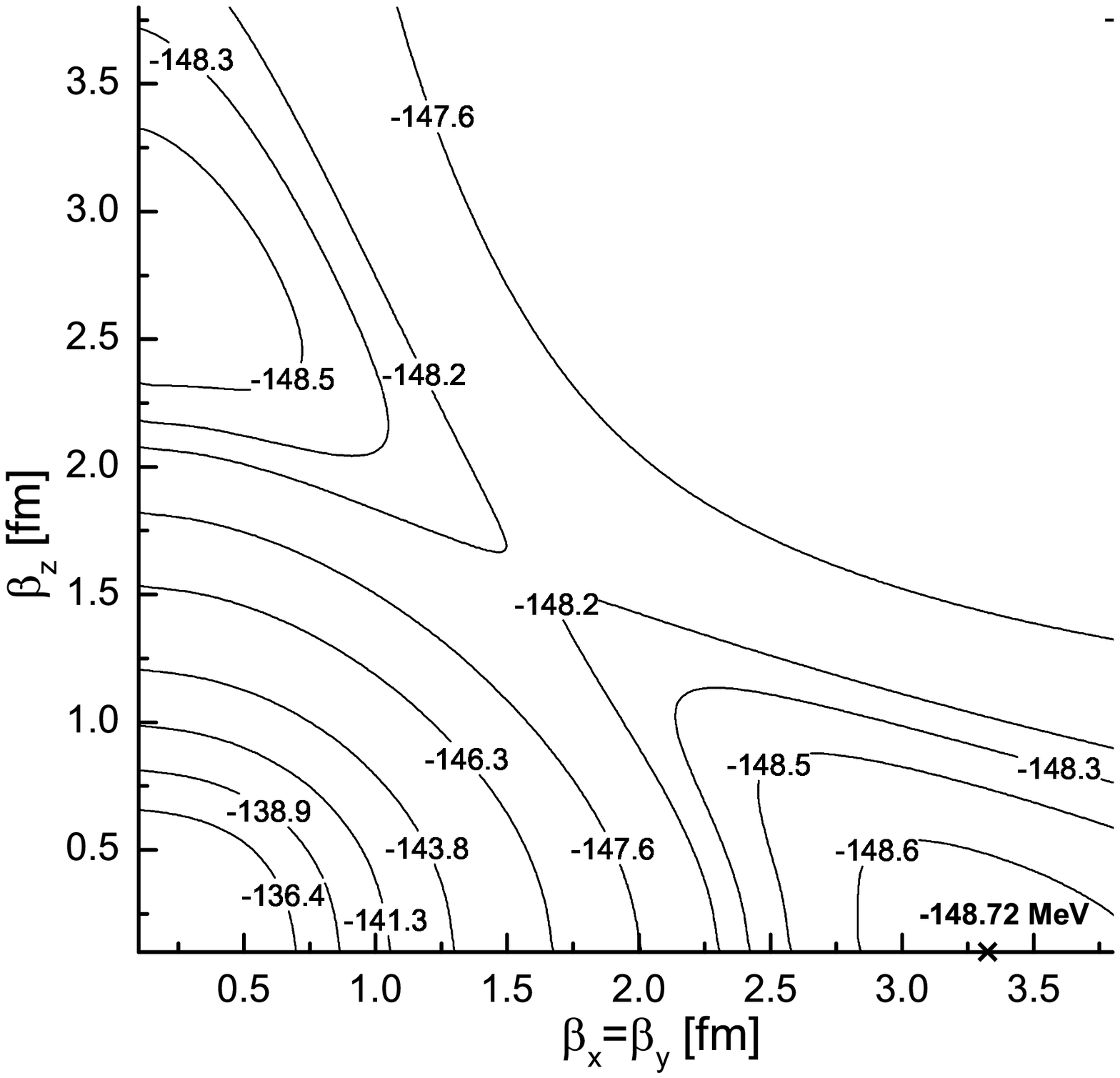}
\caption{\label{5-bxbz} Contour map of the energy surface of the  $J^\pi=5^- $ state in the two-parameter space, $\beta_{x}=\beta_{y}$ and $\beta_{z}$.}
\end{figure}
 
\begin{figure}[!h]
\centering
\includegraphics[scale=0.42]{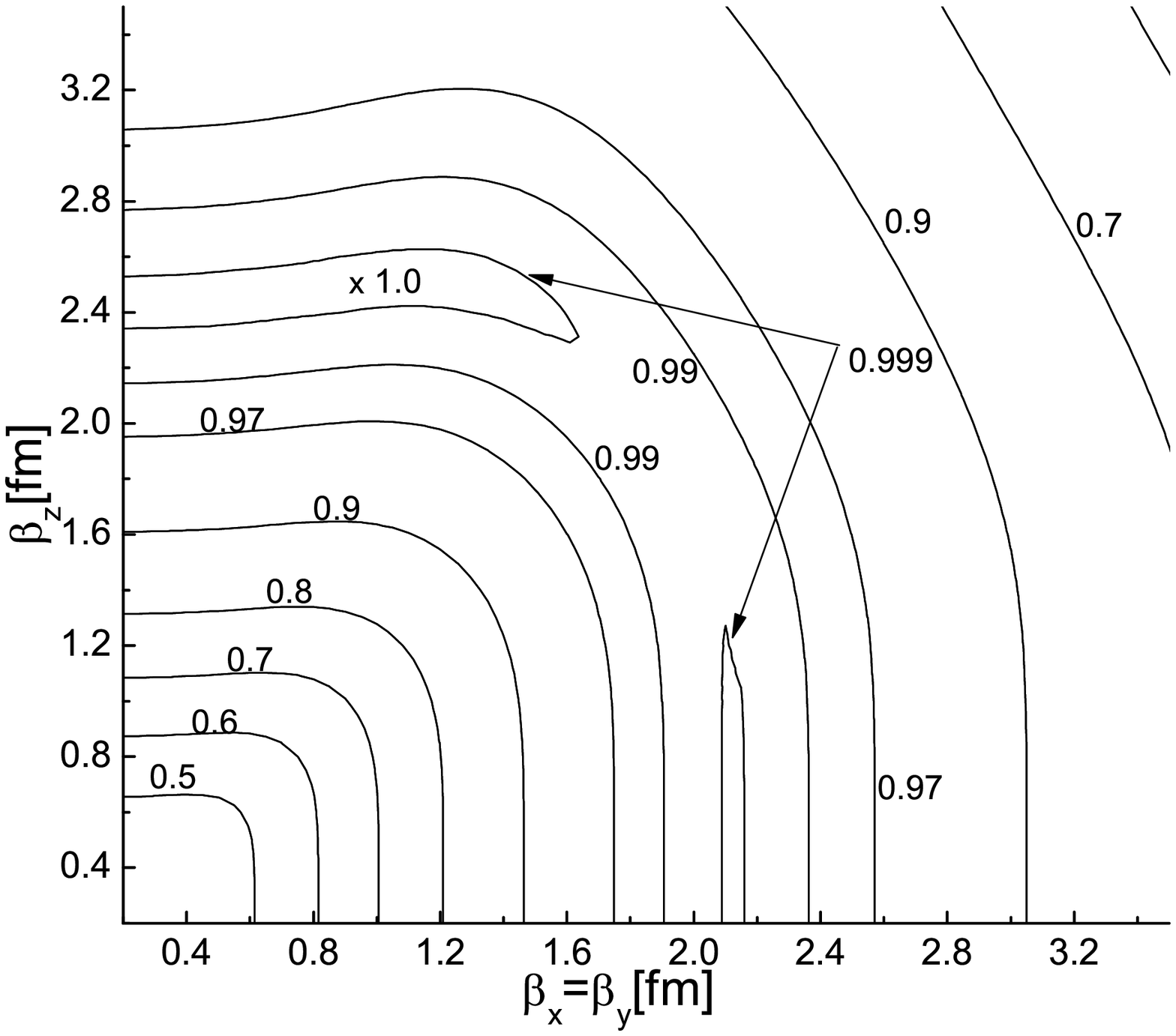}
\caption{\label{0+overlap}Contour map of the squared overlap between the $0^{+}$ wave function with $\beta_{x}=\beta_{y}=0.9$ fm, $\beta_{z} =2.5$ fm and the $0^+$ wave function with variable $\beta_{x}=\beta_{y}$ and $\beta_{z}$. Numbers attached to the contour lines are squared overlap values.}
\end{figure}

\begin{figure}[!h]
\centering
\includegraphics[scale=0.42]{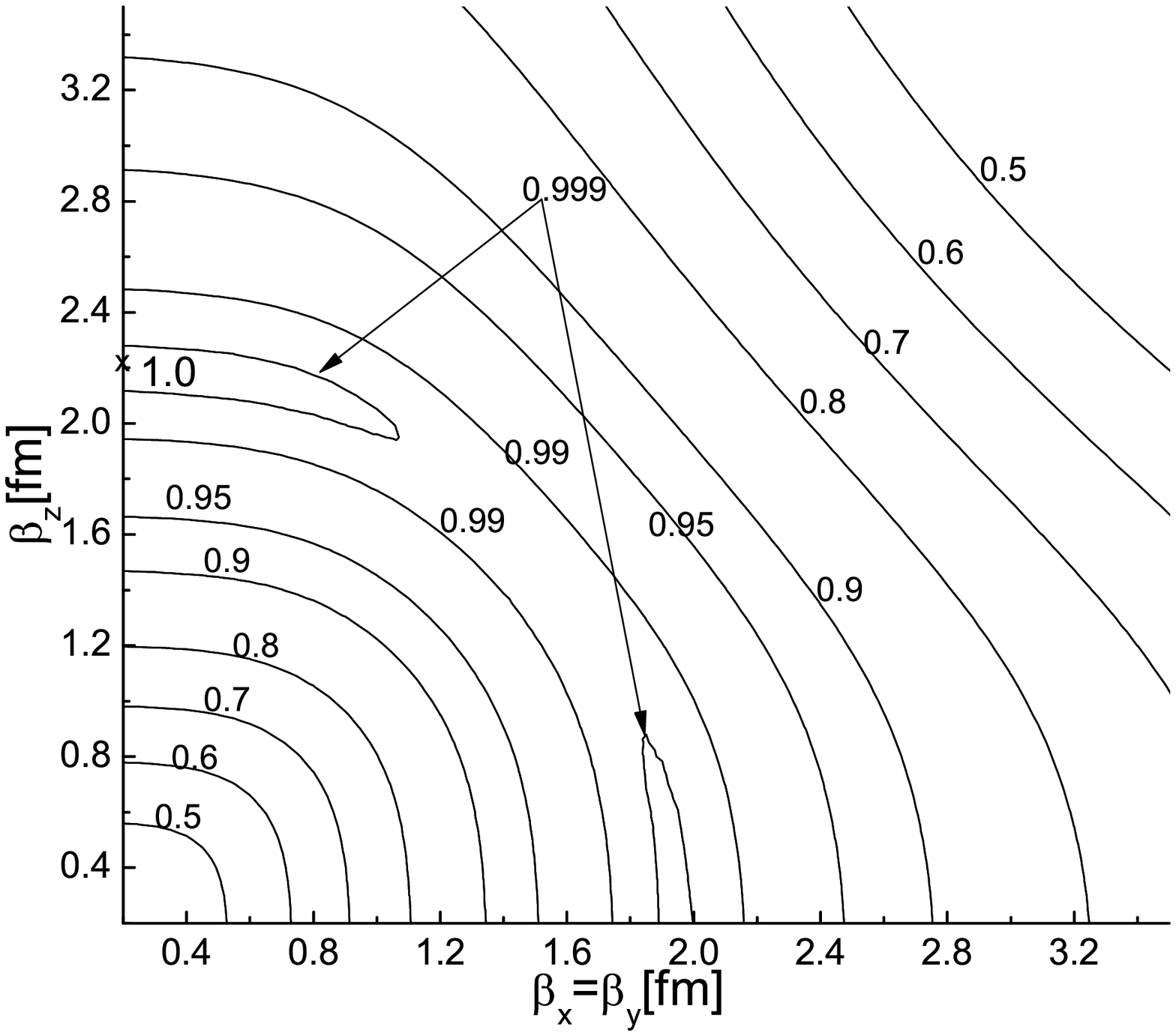}
\caption{\label{2+overlap}Contour map of the squared overlap between the $2^{+}$ wave function with $\beta_{x}=\beta_{y}=0.0$ fm, $\beta_{z} =2.2$ fm and the $2^+$ wave function with variable $\beta_{x}=\beta_{y}$ and $\beta_{z}$. Numbers attached to the contour lines are squared overlap values.}
\end{figure}

\begin{figure}[!h]
\centering
\includegraphics[scale=0.42]{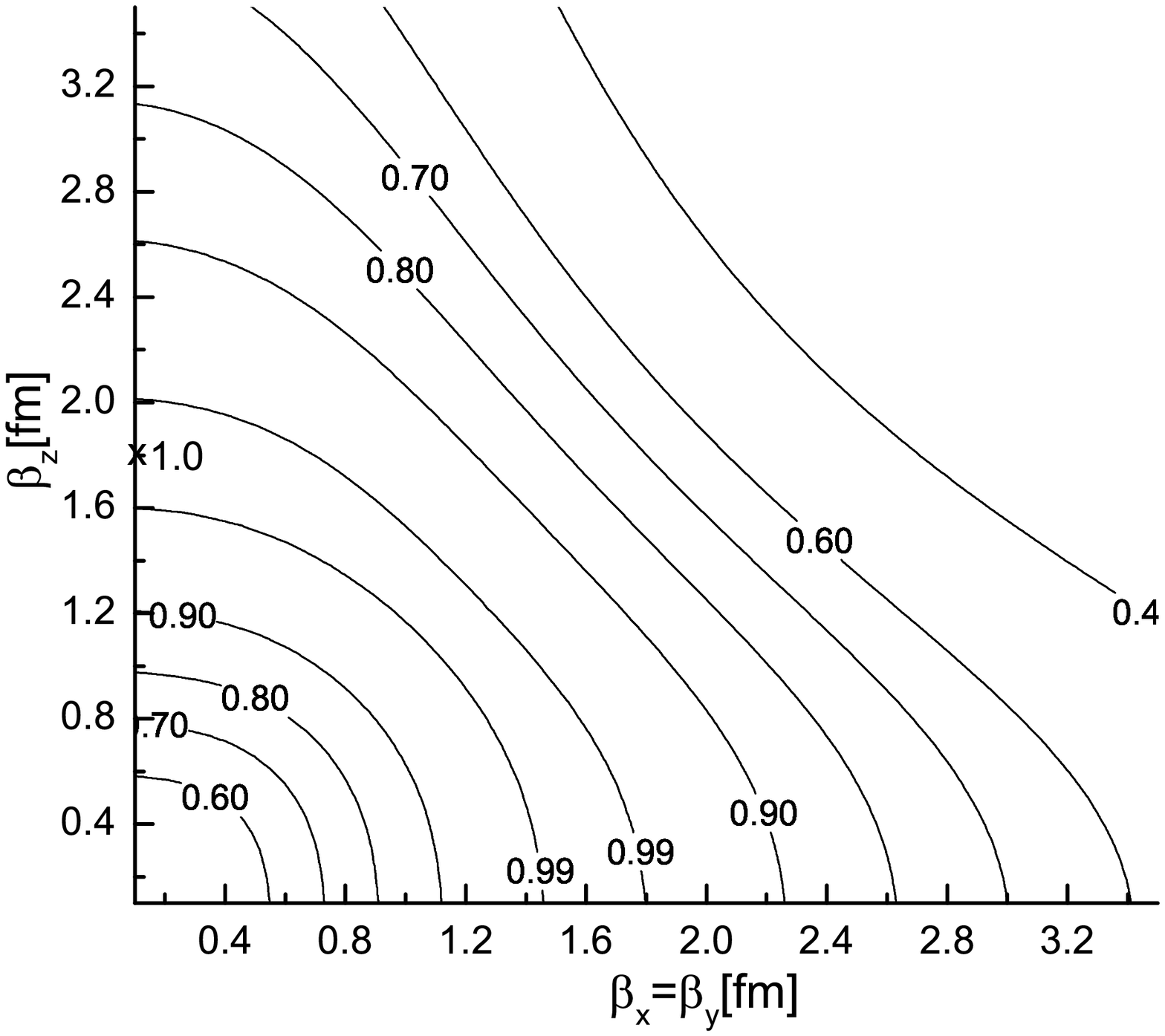}
\caption{\label{4+overlap}Contour map of the squared overlap between the $4^{+}$ wave function with $\beta_{x}=\beta_{y}=0.0$ fm, $\beta_{z} =1.8$ fm and the $4^+$ wave function with variable $\beta_{x}=\beta_{y}$ and $\beta_{z}$. Numbers attached to the contour lines are squared overlap values.}
\end{figure} 
 
\begin{figure}[!h]
\centering
\includegraphics[scale=0.42]{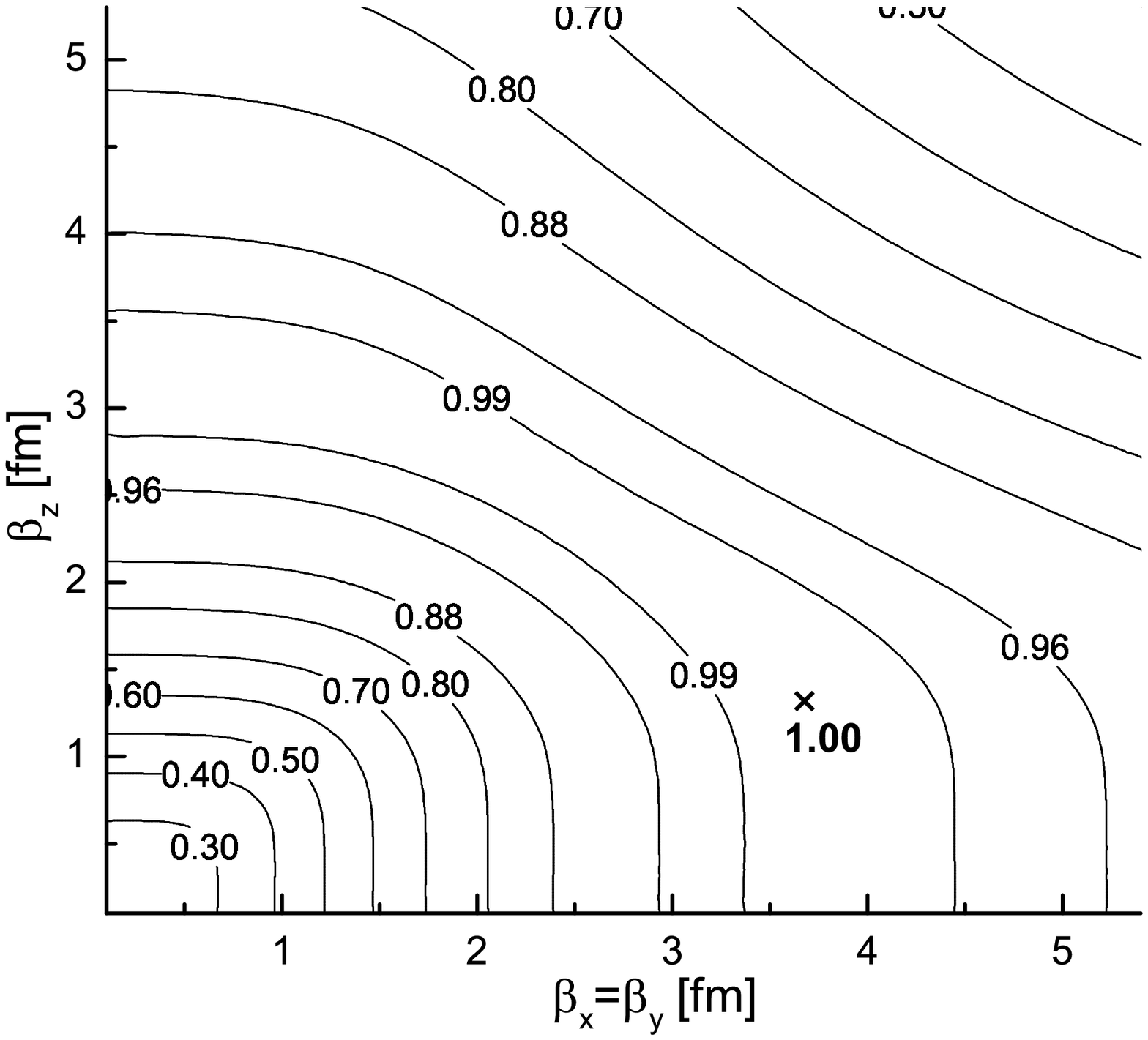}
\caption{\label{1-overlap} Contour map of the squared overlap between the $1^{-}$ wave function with $\beta_{x}=\beta_{y}=3.7$ fm, $\beta_{z} =1.4$ fm and the $1^{-}$ wave function with variable $\beta_{x}=\beta_{y}$ and $\beta_{z}$. Numbers attached to the contour lines are squared overlap values.}
\end{figure}
\begin{figure}[!h]
\centering
\includegraphics[scale=0.42]{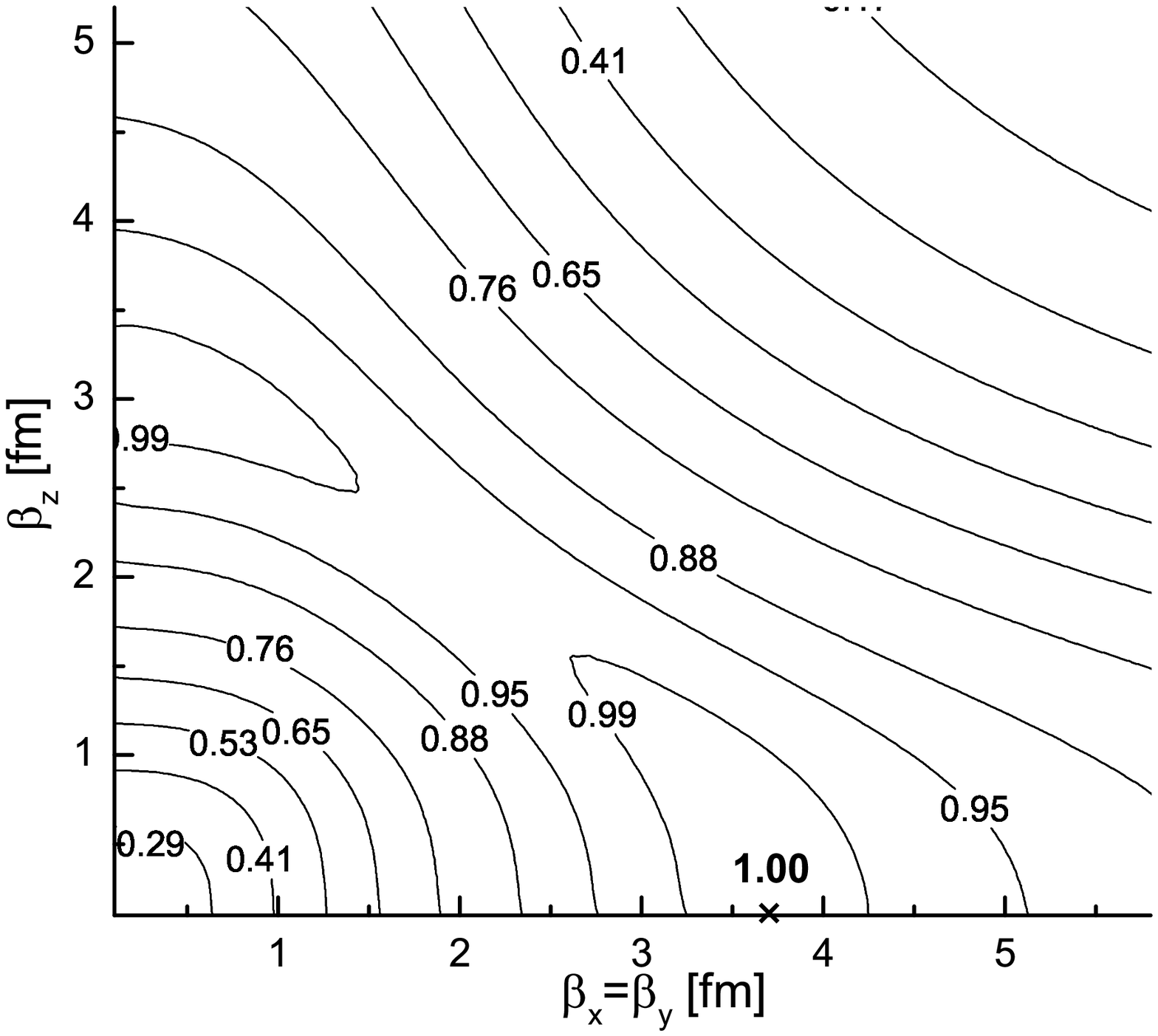}
\caption{\label{3-overlap} Contour map of the squared overlap between the $3^-$ wave function with $\beta_{x}=\beta_{y}=3.7$ fm, $\beta_{z} =0.0$ fm and the $3^-$ wave function with variable $\beta_{x}=\beta_{y}$ and $\beta_{z}$. Numbers attached to the contour lines are squared overlap values.}
\end{figure} 

\begin{figure}[!h]
\centering
\includegraphics[scale=0.42]{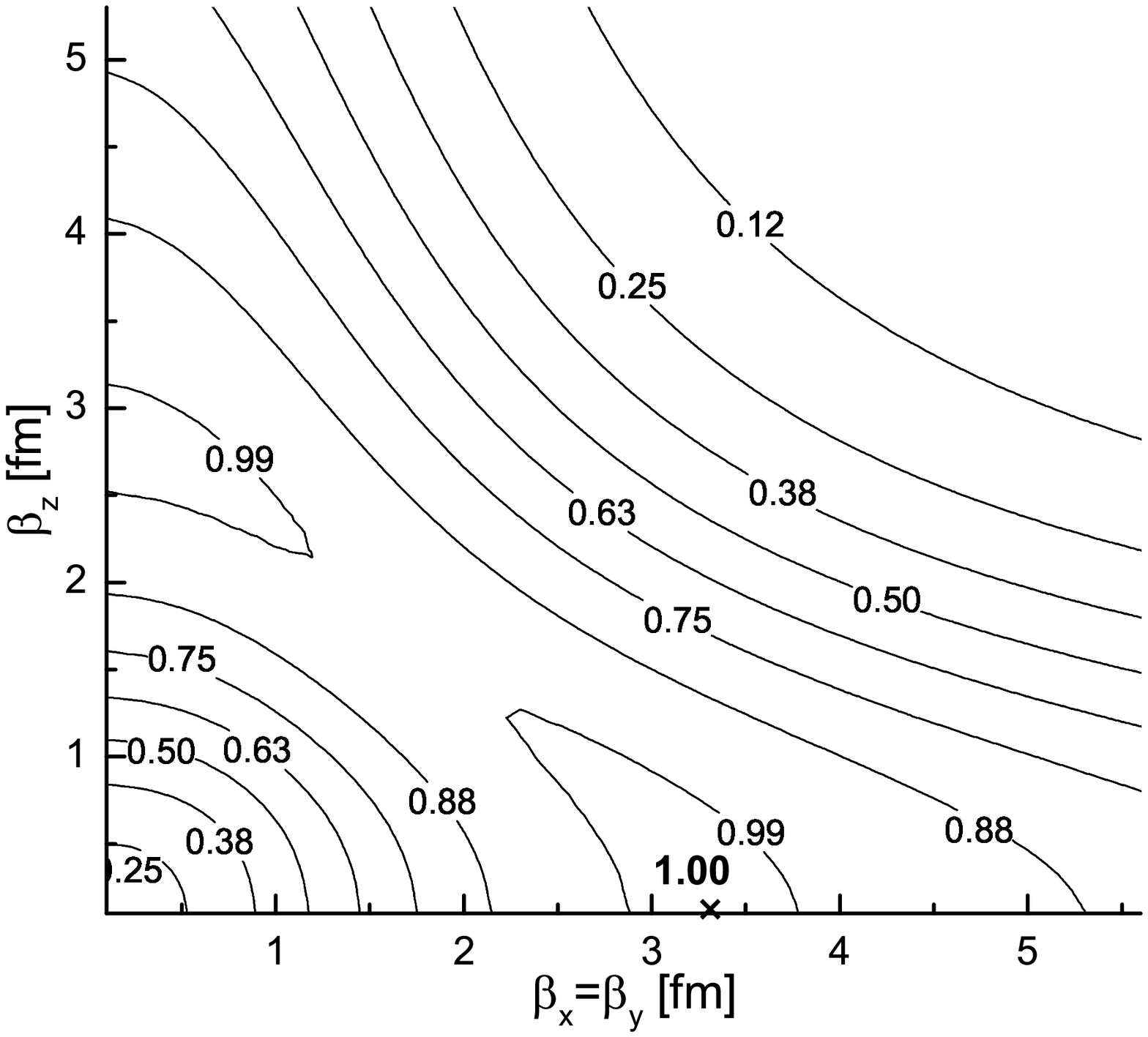}
\caption{\label{5-overlap} Contour map of the squared overlap between the $5^-$ wave function with $\beta_{x}=\beta_{y}=3.3$ fm, $\beta_{z} =0.0$ fm and the $5^-$ wave function with variable $\beta_{x}=\beta_{y}$ and $\beta_{z}$. Numbers attached to the contour lines are squared overlap values.}
\end{figure} 

In the description of ${}^{8}$Be and ${}^{12}$C  using the THSR wave function, it was found that  the  projected prolate THSR wave functions are nearly equivalent to the projected oblate THSR wave functions based on the calculations of their energy contours and the relevant squared overlaps \cite{Fu02, Fu05}.  In fact, for the general $n\alpha$ systems,  it can be demonstrated that the $n\alpha$  angular-momentum-projection THSR wave function ${\widehat \Phi}^{J^\pi}_{n\alpha}(\vect{\beta})$  obtained from a prolate intrinsic state can be obtained approximately from an oblate intrinsic state and vice versa, except for the case of strongly prolate deformation \cite{ Fu05}.  This is a very characteristic property for the THSR wave function.

In this section,  we will discuss the character of the obtained THSR-type wave function of the inversion doublet band in \nene.
Firstly, according to Eq.~(\ref{contour}),  we can obtain the contour maps of the energy surfaces of the rotational states of the inversion doublet bands in \nene\  in the two-parameter space, $\beta_{x}=\beta_{y}$ and $\beta_{z}$, namely, the energy surfaces $E^{J^\pi}(\beta_x=\beta_y, \beta_z)$ for $J^\pi=0^+, 2^+, 4^+, 1^-, 3^-, 5^-$ states. 

Figs.~\ref{0+detail} --- \ref{4+detail} show the contour maps of the energy surfaces of the  $J^\pi=0^+$, $2^+$, and  $4^+$ states of the ground-state band in \nene\ \cite{Zh12}.  It can be seen that the energy surfaces in these contour maps are rather flat. At the same time, in each contour,  there is a narrow valley connecting the prolate region and oblate region, in which the obtained binding energies vary very little.  For these positive-parity states of \nene, the minimum-energy points appear in the prolate region of the valley, which are also very close to the secondary minimum-energy points in the oblate region.  
For instance, for the energy surface of the ground state of \nene\ in Fig.~\ref{0+detail}, the energy region with $E^{0^+}(\beta_x=\beta_y, \beta_z)<-159.6$  MeV can be considered as a valley in the contour map.  In this valley, the minimum energy point  -159.85 MeV  occurs at $\beta_{x}=\beta_{y} = 0.9$ fm and $\beta_{z}=2.5$ fm in the prolate region.  And the secondary minimum energy, -159.74 MeV, appears at $\beta_{x}=\beta_{y}= 2.1$ fm and $\beta_{z}=0.0$ fm in the oblate region.  The two-minimum-energy difference  is only about 0.1 MeV despite their completely different shapes.   
 
Figs.~\ref{1-bxbz} --- \ref{5-bxbz} show the contour maps of the energy surfaces of the  $J^\pi=1^-$, $3^- $, and  $5^- $ states in the two-parameter space, $\beta_{x}=\beta_{y}$ and $\beta_{z}$, respectively. Like the positive-parity state of \nene, there is a flat valley in the contour map of the negative-parity state and the energies vary very little in this region.  It should be noted that,  different from  the positive states of the ground-state band in \nene,  the minimum points for the negative-parity states appear in the oblate regions rather than the prolate regions.  For instance, for the $J^\pi=1^-$ state in Fig.~\ref{1-bxbz} ,  the first minimum energy $-155.38$ MeV appears at $\beta_{x}=\beta_{y}=3.7 $ and $\beta_{z}=1.4$ fm  in the oblate region  while the second minimum energy $-155.37$ MeV appears at $\beta_{x}=\beta_{y}=0.7$ and $\beta_{z}=3.1$ fm  in the prolate region.  The two  minimum energies are nearly equivalent and there is a very narrow valley with a nearly flat bottom connecting the two minimum points. 

To further clarify the similarity of the projected prolate and oblate wave functions,  we will show the contour maps of the squared overlaps between the normalized projected THSR-type wave functions ${\widehat \Phi}^{J^\pi}_{\text{Ne,min}}$ with respect to the minimum energies and the corresponding normalized  projected wave functions ${\widehat \Phi}^{J^\pi}_{\text{Ne}}(\beta_x=\beta_y, \beta_z)$ with variable $\beta_{x}=\beta_{y}$ and $\beta_{z}$, namely, the following squared overlap,
\begin{equation}
\label{sqop}
O_p(\beta_x=\beta_y, \beta_z)=\frac{|\langle {\widehat \Phi}^{J^\pi}_{\text{Ne,min}}|{\widehat \Phi}^{J^\pi}_{\text{Ne}}(\beta_x=\beta_y, \beta_z)\rangle|^2}{\langle {\widehat \Phi}^{J^\pi}_{\text{Ne,min}} | {\widehat \Phi}^{J^\pi}_{\text{Ne,min}}\rangle \langle {\widehat \Phi}^{J^\pi}_{\text{Ne}}(\beta_x=\beta_y, \beta_z) | {\widehat \Phi}^{J^\pi}_{\text{Ne}}(\beta_x=\beta_y,\beta_z)\rangle}.
\end{equation} 

Figs.~\ref{0+overlap} --- \ref{4+overlap} show the contour maps for the squared overlap $O_p(\beta_x=\beta_y, \beta_z)$  for the  $J^\pi=0^+$, $2^+$, and  $4^+$ states of the ground-state band in \nene. We can see that  the projected $J^\pi$ wave function  is nearly unchanged from the optimum wave function ${\widehat \Phi}^{J^\pi}_{\text{Ne,min}}$ along the valley running from the energy minimum in the oblate region deeply into the region of prolate deformation. In other words, for the ground-state band in \nene,  the projected prolate and oblate THSR wave functions in the valley are very similar in spite of their completely different shapes.  
For instance,  Fig.~\ref{0+overlap} displays the contour map of the squared overlap between the $0^{+}$ wave function with $\beta_{x}=\beta_{y}=0.9$ fm, $\beta_{z}=2.5$ fm and the $0^{+}$ wave function with variable $\beta_{x}(=\beta_{y})$ and $\beta_{z}$.  It can be seen that oblate and prolate regions have very similar wave functions.  In particular, the obtained squared overlap between the normalized projected prolate wave function $\widehat{\Phi}_ {\text{Ne,min1}}^{0^+}$ corresponding to the state of minimum energy and the normalized projected oblate wave function $\widehat{\Phi}_ {\text{Ne,min2}}^{0^+}$ corresponding to the state of secondary minimum energy is about 0.999. This means the two wave functions with respect to their minimum points in completely different regions are almost equivalent. 

In Figs.~\ref{1-overlap} --- \ref{5-overlap}, we also show the contour maps of the squared overlaps between the normalized projected negative-parity wave functions  ${\widehat \Phi}^{J^\pi}_{\text{Ne,min}}$ with respect to the minimum energies and the corresponding  normalized projected wave functions ${\widehat \Phi}^{J^\pi}_{\text{Ne}}(\beta_x=\beta_y, \beta_z)$ with variable $\beta_{x}=\beta_{y}$ and $\beta_{z}$.  The features of these contours are very similar to the case of the positive-parity states of \nene. The obtained projected oblate THSR wave functions with respect to the minimum energies for the  $J^\pi=1^-$, $3^- $, and  $5^- $ states have very strong overlap with their respective projected prolate wave functions.
For instance, Fig.~\ref{1-overlap} displays the contour map of the squared overlap $O_p(\beta_x=\beta_y, \beta_z)$ for the  $J^\pi=1^-$ state.   It can be seen clearly that squared overlaps are more than 99$\%$  in the energetically flat region. This means the projected THSR wave functions in this valley region are very similar to one another.  At the same time, the  projected oblate  THSR wave function  with $\beta_{x}=\beta_{y}=3.7$ fm, $\beta_{z}=1.4$ fm giving the minimum energy for  the $1^{-}$ state has a squared overlap value as high as 99.98$\%$ with the $1^{-}$ wave function projected from the prolately  deformed THSR wave function with  $\beta_{x}=\beta_{y}=0.1$ fm, $\beta_{z}=3.2$ fm. Therefore,  these prolate and oblate THSR-type wave functions after angular-momentum projection are also nearly equivalent for the $J^\pi=1^-$ state of \nene. 
   
Thus, by calculating the energy surfaces  $E^{J^\pi}(\beta_x=\beta_y, \beta_z)$ and the squared overlaps  $O_p(\beta_x=\beta_y, \beta_z)$ of \nene,   
we reach the conclusion that after angular-momentum projection, the intrinsic THSR wave functions with completely different shapes become very similar, in particular,  the projected prolate THSR-type wave function of \nene\ is almost completely equivalent to the projected oblate THSR wave function and vice versa. These features are similar to the cases of  the projected THSR wave functions of ${}^{8}$Be and ${}^{12}$C studied earlier \cite{Fu02, Fu05}. Let us now try to elucidate this somewhat puzzling situation.   

\section{Nonexistence of physically oblate deformation in two-cluster systems and the meaning of oblate THSR wave function  \label{sec4}}

\subsection{Even oblate THSR wave function is of prolate character after angular-momentum projection in two-cluster systems}

As was mentioned in the previous section, after the angular-momentum projection, a prolate THSR wave function is 
almost equivalent to some oblate THSR wave function and conversely an oblate THSR wave function is almost 
equivalent to some prolate THSR wave function.   Therefore one may wonder what is the actual intrinsic deformation of the 
angular-momentum projected THSR wave function.  In the traditional description of the cluster system, 
the intrinsic state is discussed by using the Brink-GCM formalism.  In this formalism, the intrinsic state 
of any two-cluster system is necessarily prolate.  It is because any two-cluster wave function $\Phi_{L0}$ is 
expressed as follows, 
\begin{eqnarray}
 && \Phi_{L0} = {\cal A}\left\{ \chi_L(r) Y_{L0}(\widehat{r}) \phi(C_1) \phi(C_2)\right\} 
    = P^L \Phi^{\rm BGI}, \\ 
 && \Phi^{\rm BGI}  = \sum_j f_L(j) {\cal A}\left\{ \exp\left[-\gamma (\Vec{r} - {S_z}_j \Vec{e}_z)^2\right] 
    \phi(C_1) \phi(C_2)\right\}, \quad \gamma = \frac{A_1A_2}{A_1+A_2}\frac{1}{2b^2}. 
\end{eqnarray} 
Here $P^L$ is the angular-momentum projection operator and $A_k$ is the mass number of cluster 
$C_k$ ($k$=1, 2). 
The wave function $\Phi^{\rm BGI}$ is the intrinsic wave function in the Brink-GCM representation of 
$\Phi_{L0}$ and it has clearly a prolate deformation. Since the angular-momentum projected THSR wave function 
is practically equivalent to a Brink-GCM wave function $P^L \Phi^{\rm BGI}$ \cite{Fu02,Fu09,Zh13}, the Brink-GCM formalism 
may tell us that 
the former has effectively a prolate deformation even if it is obtained by the angular-momentum projection of the oblate THSR wave function.
 
Our finding that the prolate and oblate THSR wave functions can become almost equivalent after the angular-momentum 
projection makes us doubt about the above argument, because after the angular-momentum projection of $\Phi^{\rm BGI}$ 
the prolate-deformation character of $\Phi^{\rm BGI}$ may not be maintained.  In order to get rid of this doubt, 
it is desirable to judge the deformation by using not the intrinsic wave function but the angular-momentum projected 
wave function.  A good way to do such a kind of judgement, is to calculate the quadrupole moment with the angular-momentum 
projected wave function.  According to the Bohr model, the quadrupole moment $Q(L)$ of the angular-momentum $L$ 
state is related to the intrinsic quadrupole moment $Q({\rm int})$ as 
\begin{eqnarray}
  Q(L) = - \frac{L}{2L+3} Q({\rm int}). 
\end{eqnarray} 
This formula tells us that, if the deformation is prolate with positive $Q({\rm int})$, $Q(L)$ is negative, 
while $Q(L)$ is positive for oblate deformation with negative $Q({\rm int})$.  In Ref.~\cite{Ma75}, 
the calculated values of $Q(L)$, using $^{16}$O + $\alpha$ RGM, are reported showing that they are of negative 
sign for all the states of the inversion doublet bands.  This result shows, of course, that the inversion doublet 
bands are all of prolate deformation. Now, as we mentioned in Sec.~\ref{sec2}, the Brink-GCM wave functions of 
the inversion-doublet-band states are almost 100$\%$ equivalent to single THSR wave functions with 
angular-momentum projection.  Because of the equivalence of the Brink-GCM wave function with the RGM 
wave function \cite{Ho70}, we can say that the values of $Q(L)$ by the angular-momentum projected THSR wave functions are 
all of negative sign.  Thus we know that after the angular-momentum projection, not only the prolate THSR 
wave function but also the oblate THSR wave function have the character of prolate deformation. To study this question in more detail, we want to give the expression for the quadrupole moment.

When both clusters of the two-cluster system are SU(3)-scalar nuclei, such as $\alpha$, $^{16}$O, and $0s$-shell nuclei like $d$, $t$, $^3$He, the expectation value of the quadrupole moment operator calculated with an arbitrary RGM wave function $\Psi$, can be expressed 
analytically as follows 
\begin{eqnarray}
  && \langle \Psi | \frac{1}{2} \sum_i Q_{20}(i) | \Psi \rangle 
     = - \frac{L}{2L+3}\ \frac{A_1A_2}{A}\ \langle r^2 \rangle,     \label{eq:rgmqmom}  \\
  && \frac{A_1A_2}{A}\ \langle r^2 \rangle \equiv  \langle \Psi | \sum_i (\Vec{r}_i - \Vec{X}_G)^2 | 
     \Psi \rangle - (\langle R^2(C_1) \rangle + \langle R^2(C_2) \rangle),        \\
  && \langle R^2(C_k) \rangle = \langle \phi(C_k)| \sum_{i \in C_k} (\Vec{r}_i - \Vec{X}_{Gk})^2 | \phi(C_k) \rangle, 
     \ (k=1,2), \\
  && \Psi = \sqrt{\frac{A_1!A_2!}{A!}} {\cal A} \left\{ \chi_L(r)Y_{LL}(\widehat{r}) \phi(C_1) \phi(C_2)\right\},  \\ 
  && \langle \Psi | \Psi \rangle = 1, \quad \chi_L(r) = {\rm arbitrary}, \\
  && Q_{20}(i) = Q_{20}(\Vec{r}_i - \Vec{X}_G), \quad Q_{20}(\Vec{R}) = (3R_z^2 - R^2), 
\end{eqnarray}
where $\Vec{X}_G$ and $\Vec{X}_{Gk}$ stand for the center-of-mass coordinates of the total system and the 
cluster $C_k$, respectively.  A derivation of this formula of Eq.~(\ref{eq:rgmqmom}) is given in the Appendix A. 
The formula of Eq.~(\ref{eq:rgmqmom}) tells us clearly that any RGM wave function of any two-cluster system 
composed of SU(3)-scalar clusters yields a negative quadrupole moment, which is of course consistent with 
the $^{16}$O + $\alpha$ RGM results of Ref.~\cite{Ma75}. Since any THSR wave function 
after angular-momentum projection is very close to an RGM wave function \cite{Fu02,Fu03,Fu09,Zh13}, we know that any THSR wave function after angular-momentum projection yields negative quadrupole moment. Let us explain the deeper reason for this fact.

\subsection{Oblate two-cluster THSR wave function is a rotation average of a prolate THSR wave function}

\begin{figure}[htbp]
\begin{center}
\includegraphics[scale=0.35]{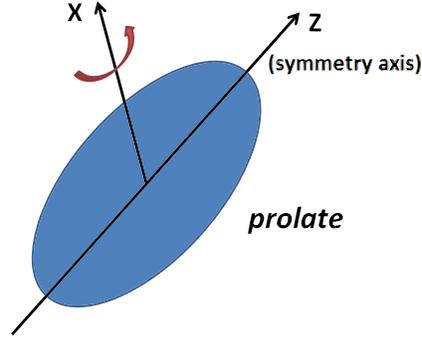}
\caption{Rotation average of a prolate THSR wave function around an axis ($x$-axis) perpendicular to the symmetry 
         axis ($z$-axis) of the prolate THSR wave function.\label{fig:1}}
\end{center}
\end{figure}

In the above we have seen that, in two-cluster systems, an oblate THSR wave function whose density distribution 
is actually oblate becomes a wave function with prolate nature after angular-momentum projection. This fact 
suggests that an oblate THSR wave function is equivalent to the rotation average of  some prolate THSR wave function. 
If we rotate a prolate THSR wave function around an axis ($x$-axis) perpendicular to the symmetry axis ($z$-axis) of 
the prolate deformation and construct a wave function by taking an average of this rotation, the density distribution 
of the rotation-average wave function will be oblate (see Fig.~\ref{fig:1}).  Let us express the rotation-average wave 
function generated from a prolate THSR wave function $\Phi^{\rm prol}(B_x=B_y,B_z)$ as 
$\Phi^{\rm AV}(B_x=B_y,B_z)$ 
\begin{eqnarray}
  \Phi^{\rm AV}(B_x=B_y,B_z) = \left(\frac{1}{2\pi} \int d\theta e^{-i\theta \ell_x}\right) 
  \Phi^{\rm prol}(B_x=B_y,B_z).  \label{eq:rotaver}
\end{eqnarray} 
We can easily prove that the wave functions obtained by the angular-momentum projection from this $\Phi^{\rm AV}$ 
are the same as those obtained by the angular-momentum projection from the original wave function $\Phi^{\rm prol}$.  
Namely even though $\Phi^{\rm AV}$ is of oblate nature, its angular-momentum projection gives the same 
wave functions as those obtained by the angular-momentum projection from the prolate wave function $\Phi^{\rm prol}$.
\begin{eqnarray}
  {\cal N}_{\rm AV} P^J_{M,0} \Phi^{\rm AV}(B_x=B_y,B_z) = {\cal N}_{\rm prol} P^J_{M,0} \Phi^{\rm prol}(B_x=B_y,B_z), 
\end{eqnarray} 
where $P^J_{M,0}$ is the angular-momentum projection operator, and ${\cal N}_{\rm AV}$ and ${\cal N}_{\rm prol}$ are normalization constants.  

\begin{figure}[htbp]
\begin{center}
\includegraphics[scale=0.4]{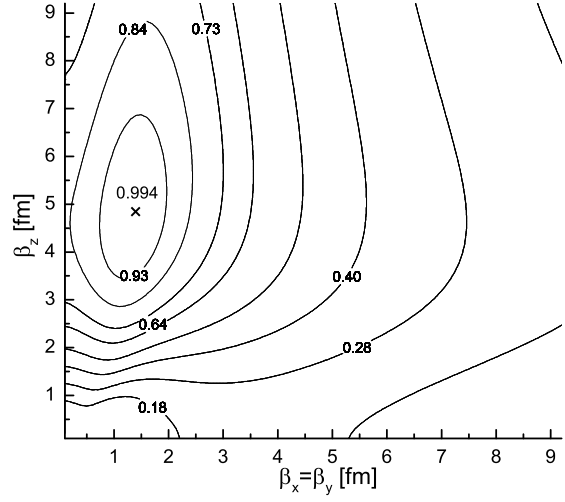}
\caption{Squared overlap $|O(B_x=B_y, B_z)|^2$ of the rotation-average wave function $\Phi^{\rm AV}(B_x = B_y, B_z)$ with the oblate $^{16}$O + $\alpha$ THSR wave function $\Phi^{\rm obl}(\tilde{B}_x, \tilde{B}_y = \tilde{B}_z)$ which gives the minimum energy of the $J^\pi = 1^-$ state after the angular-momentum projection.  The values of 
$\tilde{B}_k$ are $(\tilde{\beta}_x, \tilde{\beta}_y, \tilde{\beta}_z) = (1.4\ {\rm fm}, 3.7\ {\rm fm}, 3.7\ {\rm fm})$, 
where $\tilde{B}_k^2 = b^2 + 2\tilde{\beta}_k^2$.\label{fig:2}}
\end{center}
\end{figure}

In order to understand this point, we will prove, up to the first order of the deformation parameter $(B_x - B_z)$, that the rotation average of a 
prolate THSR wave function $\Phi(B_x=B_y, B_z)$ becomes an oblate THSR wave function:     
\begin{eqnarray}
  && \Phi(B_x=B_y,B_z) = {\cal A} \left\{ \exp \left[-\gamma_x (r_x^2 + r_y^2) - \gamma_z r_z^2\right] 
                   \phi(C_1) \phi(C_2) \right\},  \\
  && \hspace{2cm} \gamma_k = \left(\frac{A_1A_2}{A_1+A_2}\right) \frac{1}{2B_k^2} \quad (k = x, y, z), \\
  && \exp \left[-\gamma_x (r_x^2 + r_y^2) - \gamma_z r_z^2\right] 
     = \exp \left[-\left( \frac{2}{3}\gamma_x + \frac{1}{3}\gamma_z \right) \Vec{r}^2 \right]  \nonumber \\
  && \hspace{2cm} \times \left\{ 1-\left(\frac{1}{3}\gamma_z - \frac{1}{3}\gamma_x\right)
      \sqrt{\frac{16\pi}{5}}r^2 Y_{20}(\widehat{r}) + \cdots \right\}    \\
  && \left(\frac{1}{2\pi} \int d\theta e^{-i\theta \ell_x}\right) \exp \left[-\gamma_x (r_x^2 + r_y^2) - 
     \gamma_z r_z^2\right] = \exp \left[-\left( \frac{2}{3}\gamma_x + \frac{1}{3}\gamma_z \right) \Vec{r}^2 \right] 
          \nonumber \\
  && \hspace{2cm} \times \left\{ 1-\left(\frac{1}{3}\gamma_z - \frac{1}{3}\gamma_x\right)
     \sqrt{\frac{16\pi}{5}}r^2 \left(\frac{1}{2\pi} \int d\theta e^{-i\theta \ell_x}\right) Y_{20}(\widehat{r}) 
     + \cdots \right\}  \\
  && \hspace{1cm} = e^{i(\pi/2) \ell_y} \exp \left[-\left( \frac{2}{3}\gamma_x + \frac{1}{3}\gamma_z \right) 
     \Vec{r}^2 \right]  \nonumber \\
  && \hspace{2cm} \times \left\{ 1 + \frac{1}{2} \left(\frac{1}{3}\gamma_z - \frac{1}{3}\gamma_x\right)
     \sqrt{\frac{16\pi}{5}}r^2 Y_{20}(\widehat{r}) + \cdots \right\}     \\
  && \hspace{1cm} \approx  \exp \left[ -\gamma'_x r_x^2 -\gamma'_y (r_y^2 + r_z^2)\right]   \\
  && \hspace{2cm}  \gamma'_x = \gamma_x, \quad \gamma'_y = \gamma'_z = \frac{1}{2} (\gamma_x + \gamma_z),              
\end{eqnarray} 
where use is made of the following relations 
\begin{eqnarray}
  && e^{-i\theta \ell_x} = e^{i(\pi/2) \ell_y} e^{i\theta \ell_z} e^{-i(\pi/2) \ell_y},     \\
  && e^{-i\theta \ell_x}Y_{20} = e^{i(\pi/2) \ell_y} \sum_M d^2_{M0}(\pi/2)  e^{i\theta M} Y_{2M},   \\
  && \left(\frac{1}{2\pi} \int d\theta e^{-i\theta \ell_x}\right) Y_{20} =  -\frac{1}{2}  e^{i(\pi/2) \ell_y} Y_{20}. 
\end{eqnarray} 
We thus have, up to the first order of the deformation parameter $(B_x - B_z)$, 
\begin{eqnarray}
  &&\left(\frac{1}{2\pi} \int d\theta e^{-i\theta \ell_x}\right) \Phi(B_x=B_y, B_z) \approx \Phi(B'_x, B'_y=B'_z), 
    \label{eq:rotavformA}  \\
  &&\gamma'_x = \gamma_x, \quad \gamma'_y = \gamma'_z = \frac{1}{2} (\gamma_x + \gamma_z),  \quad 
    \gamma'_k = \left(\frac{A_1A_2}{A_1+A_2}\right) \frac{1}{2(B'_k)^2}\quad  (k = x, y, z).   \label{eq:arotaver}
\end{eqnarray} 
From the relation $\gamma_x = \gamma_y > \gamma_z$ $(B_x = B_y < B_z)$, we have $\gamma'_x > \gamma'_y = \gamma'_z$  
$(B'_x < B'_y = B'_z)$ which means that the rotation average of a prolate THSR wave function $\Phi(B_x=B_y < B_z)$ is 
approximately an oblate THSR wave function $\Phi(B'_x < B'_y=B'_z)$, up to the first order of the 
deformation parameter $(B_x - B_z)$.   

Now we study numerically, not up to the first order of the deformation parameter $(B_x - B_z)$ but up to all orders, how  correct it is to say that an oblate THSR wave function 
$\Phi^{\rm obl}(\tilde{B}_x, \tilde{B}_y = \tilde{B}_z)$ is equivalent to the rotation-average wave function 
$\Phi^{\rm AV}(B_x = B_y, B_z)$ constructed from some prolate THSR wave function $\Phi^{\rm prol}(B_x = B_y, B_z)$.   
The construction of the rotation-average wave function is obtained from Eq.~(\ref{eq:rotaver}). 
For this purpose we calculate the overlap $O(B_x, B_z)$ between the normalized oblate THSR wave function of 
$\Phi^{\rm obl}(\tilde{B}_x, \tilde{B}_y = \tilde{B}_z)$ and the normalized rotation-average wave function of 
$\Phi^{\rm AV}(B_x = B_y, B_z)$  with various values of $(B_x=B_y, B_z)$:  
\begin{eqnarray}
  && O(B_x=B_y, B_z) = \tilde{O}(\tilde{B}_x, \tilde{B}_y=\tilde{B}_z; B_x=B_y, B_z)  \\
  && \hspace{1cm} = {\cal N}\ \langle \Phi^{\rm obl}(\tilde{B}_x, \tilde{B}_y = \tilde{B}_z) | 
     \Phi^{\rm AV}(B_x = B_y, B_z) \rangle, \\
  && {\cal N} = \left( ||\Phi^{\rm obl}(\tilde{B}_x, \tilde{B}_y = \tilde{B}_z)|| \cdot 
     ||\Phi^{\rm AV}(B_x=B_y, B_z)|| \right)^{-1},  \\
  && ||\Psi|| = \sqrt{\langle \Psi | \Psi \rangle}.  
\end{eqnarray}
Let us first discuss the odd parity states.
In Fig.~\ref{fig:2} we give the contour map of the squared overlap $|O(B_x=B_y, B_z)|^2$  in the plane of 
$(\beta_x=\beta_y, \beta_z)$ where $B_k^2 = b^2 + 2\beta_k^2$~in the case of the oblate THSR wave function 
$\Phi^{\rm obl}(\tilde{B}_x, \tilde{B}_y = \tilde{B}_z)$ which gives the minimum energy of the $J^\pi = 1^-$ state 
after the angular-momentum projection.  The values of $\tilde{B}_k$ are $(\tilde{\beta}_x, \tilde{\beta}_y, 
\tilde{\beta}_z) = (1.4\ {\rm fm}, 3.7\ {\rm fm}, 3.7\ {\rm fm})$, where $\tilde{B}_k^2 = b^2 + 2\tilde{\beta}_k^2$. 
We see in this figure that the squared overlap can surely become almost unity for an initially prolate THSR wave function 
$\Phi^{\rm prol}(B_x = B_y, B_z)$  with $\beta_x = \beta_y \approx 1.3$ fm, $\beta_z \approx$ 4.7 fm. 
Similarly, we have confirmed that the oblate THSR wave functions which give the minimum-energies of the 
$J^\pi = 3^-$ and $J^\pi = 5^-$ states after the angular-momentum projection are almost 100$\%$ equivalent to the 
rotation-average wave functions of some prolate THSR wave functions. 

\begin{figure}[htbp]
\begin{center}
\includegraphics[scale=0.42]{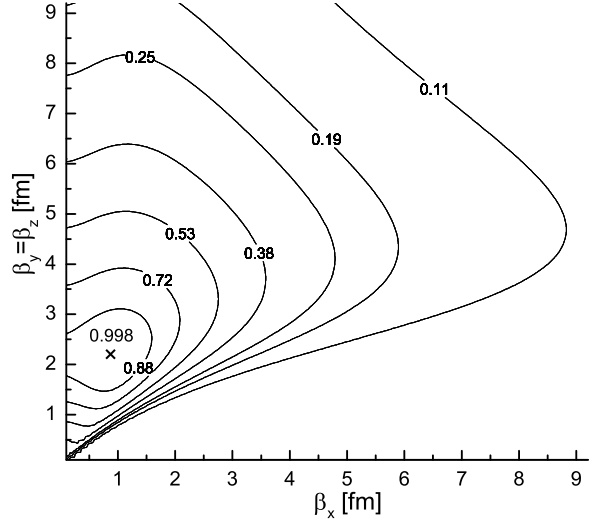}
\caption{Squared overlap $|\widehat{O}(B_x, B_y=B_z)|^2$ of the oblate $^{16}$O + $\alpha$ THSR wave function 
$\Phi^{\rm obl}(B_x, B_y = B_z)$ with the rotation-average wave function 
$\Phi^{\rm AV}(\tilde{B}_x = \tilde{B}_y, \tilde{B}_z)$ obtained from the prolate wave function 
$\Phi^{\rm prol}(\tilde{B}_x=\tilde{B}_y, \tilde{B}_z)$ which gives the minimum energy of the $J^\pi = 0^+$ 
ground state after the angular-momentum projection. The values of $\tilde{B}_k$ are
$(\tilde{\beta}_x, \tilde{\beta}_y, \tilde{\beta}_z) = (0.9\ {\rm fm}, 0.9\ {\rm fm}, 2.5\ {\rm fm})$, 
where $\tilde{B}_k^2 = b^2 + 2\tilde{\beta}_k^2$.\label{fig:3}}
\end{center}
\end{figure}

Let us now discuss the ground state $J^{\pi}=0^{+}$.
As we showed in Ref.~\cite{Zh12}, for each prolate THSR wave function of the ground-state band member state of $^{20}$Ne, there exists an oblate THSR wave function which is almost 100$\%$ equivalent to the angular-momentum projected prolate THSR wave function. We can guess that such oblate THSR wave function is almost equivalent to the rotation average of the prolate wave function.  In Fig.~\ref{fig:3} we give the contour map of the squared overlap 
$|\widehat{O}(B_x, B_y=B_z)|^2$  in the plane of $(\beta_x, \beta_y=\beta_z)$ in the case of the prolate THSR wave 
function $\Phi^{\rm prol}(\tilde{B}_x=\tilde{B}_y, \tilde{B}_z)$ which gives the minimum energy of the $J^\pi = 0^+$ ground state after the angular-momentum projection.  The values of $\tilde{B}_k$ are $(\tilde{\beta}_x, \tilde{\beta}_y, \tilde{\beta}_z) = (0.9\ {\rm fm}, 0.9\ {\rm fm}, 2.5\ {\rm fm})$.  Here $\widehat{O}(B_x, B_y=B_z)$ is defined as 
\begin{eqnarray}
 \widehat{O}(B_x, B_y=B_z) = \tilde{O}(B_x, B_y= B_z; \tilde{B}_x=\tilde{B}_y, \tilde{B}_z). 
\end{eqnarray}
We see the maximum value of $|\widehat{O}(B_x, B_y=B_z)|^2$ is almost unity around the point with $\beta_x \approx$ 0.9 fm and $\beta_y=\beta_z \approx$ 2.1 fm where $\Phi^{\rm prol}(\tilde{B}_x=\tilde{B}_y, \tilde{B}_z)$ and 
$\Phi^{\rm obl}(B_x, B_y = B_z)$ were found to be almost equivalent after angular-momentum projection in 
Ref.~\cite{Zh12}. 

\begin{figure}[htbp]
\begin{center}
\includegraphics[scale=0.75]{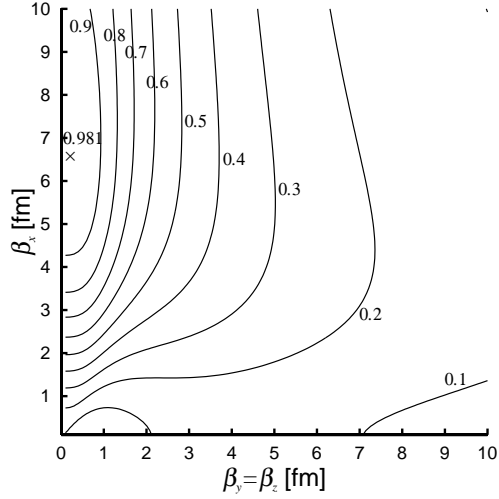}
\caption{Contour map of the squared overlap between the $\alpha$ - $\alpha$ oblate THSR wave function with 
$\beta_x$ = 0.1 fm, $\beta_y$ = $\beta_z$ = 4.4 fm and the rotation-average wave function constructed from the 
THSR wave function with various $\beta_x$, $\beta_y$ = $\beta_z$.\label{fig:3b}}
\end{center}
\end{figure}

In Ref.~\cite{Fu02} it is reported that the $J^\pi = 0^+$ $\alpha$ - $\alpha$ wave function $\Phi^{\rm obl}_{0^+}$ 
projected from the oblate THSR wave function $\Phi^{\rm obl}$ around $\beta_x$ = 0.1 fm, 
$\beta_y$ = $\beta_z$ = 4.4 fm has almost the same energy within 50 keV as the minimum energy given 
by the $J^\pi = 0^+$ wave function $\Phi^{\rm prol.A}_{0^+}$ projected from the prolate THSR $\Phi^{\rm prol.A}$ 
with $\beta_x = \beta_y$ = 1.8 fm, $\beta_z$ = 7.8 fm.  $\Phi^{\rm obl}_{0^+}$ is almost the same as 
$\Phi^{\rm prol.A}_{0^+}$ with the squared overlap 
$|\langle \Phi^{\rm obl}_{0^+} |\Phi^{\rm prol.A}_{0^+} \rangle|^2 =0.99$.   $\Phi^{\rm prol.A}_{0^+}$ is also 
almost equivalent with the wave functions projected from rather wide region of prolate THSR wave functions. 
For example the wave function $\Phi^{\rm prol.B}_{0^+}$ projected from the prolate THSR $\Phi^{\rm prol.B}$ 
with $\beta_x = \beta_y$ = 0.1 fm, $\beta_z$ = 6.6 fm has the squared overlap of almost unity with 
$\Phi^{\rm prol.A}_{0^+}$, $|\langle \Phi^{\rm prol.B}_{0^+} |\Phi^{\rm prol.A}_{0^+} \rangle|^2 =0.99$. 
In Fig.~\ref{fig:3b} we show that the oblate THSR wave function $\Phi^{\rm obl}$ is almost equivalent to 
the rotation-average wave function $\Phi^{\rm AV}(\rm prol.B)$ constructed from the prolate THSR wave function 
$\Phi^{\rm prol.B}$ with the squared overlap of almost unity, 
$|\langle \Phi^{\rm AV}(\rm prol.B) |\Phi^{\rm obl} \rangle|^2 =0.98$.  Of course, all the above discussion confirms our physical intuition displayed in Fig.~\ref{fig:1}.

In conclusion, we now understand why an angular momentum projected prolate or {\it oblate} THSR wave function gives practically the same energy, the latter nonetheless having prolate character intrinsically. Namely, e.g., the oblate minimum in Fig.~\ref{0+detail} can be considered as corresponding to a rotation around an axis perpendicular to the long symmetry axis, see Fig.~\ref{fig:1}. Additionally, it so happens that the ground state of \nene\ is such a stable prolate rotor that turning it like in Fig.~\ref{fig:1} does practically not bring any gain nor loss of energy. On the other hand the THSR wave function contains already so much of quantum fluctuations with respect to a pure Slater determinant that angular-momentum projection does bring almost no gain in energy whatsoever. This can be seen, for example, at the spherical point in Fig.~\ref{0+detail} with an energy loss of only 250 keV with respect to the absolute minimum.

\subsection{Prolate 3$\alpha$ THSR wave function is a rotation average of an oblate THSR wave function}

\begin{figure}[htbp]
\begin{center}
\includegraphics[scale=0.7]{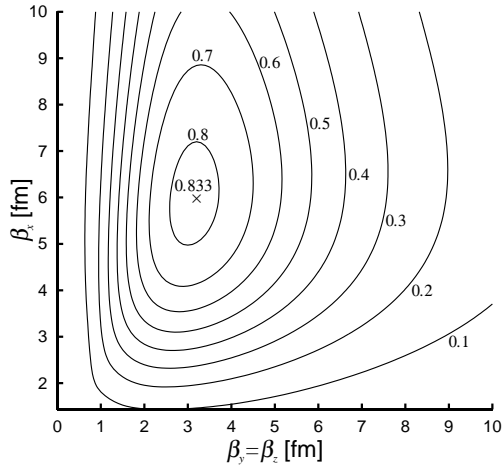}
\caption{The contour map of the squared-overlap values of the rotation-average wave function obtained from the 
oblate $3\alpha$ THSR wave function with $\beta_x = \beta_y$ = 5.7 and $\beta_z$ = 1.3 with $3\alpha$ THSR wave 
functions with various $\beta_x$ and $\beta_y = \beta_z$.\label{fig:7} }
\end{center}
\end{figure}

A further remarkable investigation of this work concerns the following. For the three $\alpha$ system,
the THSR wave function has the puzzling feature that, after angular-momentum projection, the prolate THSR 
wave function is almost equivalent to an oblate THSR wave function and also to a spherical THSR wave function.  
In the case of two-cluster systems, it was just clarified that the oblate THSR wave function is 
almost equivalent to the rotation average of some prolate THSR wave function around an axis perpendicular to 
the symmetry axis of the prolate deformation.  It implies that the oblate deformation is not the physical deformation 
of the two-cluster system.  This is assured by the fact that the quadrupole moment by the angular-momentum 
projected wave function generated from the oblate THSR wave function has the negative sign which means that the 
intrinsic deformation is prolate.   Since the intrinsic wave function is the instantaneous (or adiabatic) wave function 
of the rotating system, it is natural that the intrinsic wave function of the two-cluster system has prolate 
deformation.  While the oblate THSR wave function is interpreted as the rotation average of the prolate THSR wave 
function around an axis, the spherical THSR wave function can be interpreted as the three-dimensional rotation average 
of the prolate THSR wave function. In any case all three projected THSR wave function, be it  with intrinsic prolate, oblate, or spherical shapes yield almost degenerate energies. This only means that the THSR wave function already contains so strong quantum fluctuations that an additional projection does not bring a noticeable gain in energy.  

On the other hand, the puzzle that, after angular-momentum projection, the prolate THSR 
wave function is almost equivalent to an oblate THSR wave function has also been reported in the 3$\alpha$ 
system~\cite{Fu05}.  The calculated result of the quadrupole moment of the first $2^+$ state of $^{12}$C by the 
3$\alpha$ THSR wave function gives us the positive sign indicating the oblate deformation of the intrinsic state 
of this state.  The positive sign of the quadrupole moment of the first $2^+$ state of $^{12}$C was also reported 
in the 3$\alpha$ Brink-GCM calculation of Ref.~\cite{De87}.  Since the 3$\alpha$ clusters lie, considered at a given instant of time, 
in a plane, it is natural that the instantaneous (or adiabatic) wave function of the 3$\alpha$ rotating system
which is the intrinsic wave function has oblate deformation.  The existence of the prolate THSR wave function 
which is almost equivalent to an oblate THSR wave function after angular-momentum projection can be explained, 
at least in the first order approximation of the deformation parameter ($B_x - B_z$), by the idea of the rotation 
average of the oblate THSR wave function around an axis perpendicular to the symmetry axis of the oblate deformation: 
\begin{eqnarray}
  &&\left(\frac{1}{2\pi} \int d\theta e^{-i\theta \ell_x}\right) \Phi_{3\alpha}(B_x=B_y, B_z) \approx 
    \Phi_{3\alpha}(B'_x, B'_y=B'_z),     \label{eq:rotavformB}  \\
  &&\Phi_{3\alpha}(C_x, C_y, C_z) = {\cal A} \left[ \exp \left\{ -2\sum_{j=1}^3 \sum_{k=x,y,z} 
    \left(\frac{(X_{jk} - X_{Gk})^2}{C_k^2}\right)\right\} \prod_{j=1}^3 \phi(\alpha_j) \right], \\
  &&\frac{1}{{B'_x}^2} = \frac{1}{B_x^2}, \quad \frac{1}{{B'_y}^2} = \frac{1}{{B'_z}^2} = 
    \frac{1}{2} \left( \frac{1}{B_x^2} + \frac{1}{B_z^2} \right).    
\end{eqnarray}  
We can prove the relation of Eq.~(\ref{eq:rotavformB}) just in the same way as we have proven the relation of 
Eq.~(\ref{eq:rotavformA}). 
Since there holds $(1/{B'_x}^2) - (1/{B'_z}^2)$ = $(1/2) [(1/{B_z}^2) - (1/{B_x}^2)]$, we see that from the 
oblate deformation of $B_x=B_y > B_z$, we obtain the prolate deformation of $B'_x > B'_y = B'_z$. 
We here should note that the rotation average is made not for the density distribution which is
positive-valued but for the wave function which takes both positive and negative values.
We have studied numerically, not only up to the first order of the deformation parameter $(B_x - B_z)$ but up to all orders that the statement that the prolate THSR wave function being practically equivalent to an oblate THSR wave function is absolutely correct.  In Ref.~\cite{Fu05} it is shown that the $J^\pi = 0^+$ 
wave function obtained from the oblate THSR wave function with $\beta_x = \beta_y$ = 5.7 and $\beta_z$ = 1.3 has a large squared overlap with a value greater than 0.95 with the $J^\pi = 0^+$ wave functions obtained from the prolate THSR wave 
functions with $\beta_x = \beta_y \approx$ 3 and $\beta_z \approx$ 6.5.  In Fig.~\ref{fig:7} 
we show the contour map of the squared-overlap values of the rotation-average wave function obtained from the 
oblate THSR wave function with $\beta_x = \beta_y$ = 5.7 and $\beta_z$ = 1.3 with THSR wave functions with 
various $\beta_x$ and $\beta_y = \beta_z$.  We see that the squared overlap values are surely large for 
$\beta_y = \beta_z \approx$ 3 and $\beta_x \approx$ 6.5.

\section{Container picture of nuclear cluster dynamics and nuclear molecular structure due to the inter-cluster Pauli repulsion \label{sec5}}

 Clusters in the THSR wave function in low density systems make mutually independent nonlocalized motion occupying the lowest orbit of the 
 harmonic-oscillator-like mean-field potential of clusters characterized by the size parameter $B$ whose magnitude is 
 similar to the radius of the system.   In systems of 3$\alpha$'s~\cite{To01,Fu03,Fu05} and 4$\alpha$'s~\cite{To01,Fu10} 
 we know that the excitation mode of the system is well described by the Hill-Wheeler equation of the size parameter $B$ 
 treated as the generator coordinate.  Therefore we see that the excitation of the system is described firstly by 
 the dynamics of the size parameter $B$ which is adopted as the generator coordinate and secondly by the excitation 
 of the single-particle motion of clusters in the cluster mean-field potential.   We will call our new understanding of 
 nuclear cluster dynamics the container picture of nuclear clustering, by which we aim to stress that the central 
 quantity of cluster dynamics is the size parameter $B$ of the self-consistent mean-field potential of clusters 
 which we call the container.  The name ``container picture" may sound more appropriate for three (or more)-cluster systems because, for example, in the 3$\alpha$ system it describes the ground state (small $B$-parameter) and 3$\alpha$-gas states (large $B$-parameter) on the same 
 footing.  In this container picture the existence of cluster-gas states is natural and the formation mechanism of 
 cluster-gas states is just the spatial expansion of the container ($B$-parameter going from small to large).  When we compare the container picture of cluster 
 dynamics with the traditional description of cluster dynamics which uses explicitly wave functions with 
 inter-cluster separation coordinates, the new understanding corresponds to a collective-motion picture 
 characterized by the size parameter $B$. When $B$ has obtained a large value the clusters become more or less independent. They have, however, to respect the excluded volume (see Sec.~\ref{intro}) which is due to the Pauli principle, what leads to scattering processes among the clusters. It is in this way that, e.g., the $\alpha$ condensate is depleted by about 30$\%$ in the Hoyle state of \cc~\cite{Ya05}. 

Now we explain how the idea of the parity-violating deformation of localized $^{16}$O + $\alpha$ clustering for 
the inversion-doublet bands of $^{20}$Ne can be justified in this container picture of cluster dynamics which 
assumes nonlocalized clusters.  The parity-violating deformation is a property of the intrinsic state 
which is the instantaneous (or adiabatic) quantum state of the rotation of the nucleus. 
Since the instantaneous configuration of two clusters is of prolate shape, the prolate THSR wave function is the 
intrinsic state of the system and the oblate THSR wave function is not the intrinsic state but rather a mathematical 
object which expresses the rotation-average of the intrinsic state.  The spherical THSR wave function expresses the 
time average of the fully three-dimensional rotational motion, namely the angular-momentum projected state of the intrinsic state 
(the prolate THSR wave function). We, however, also need to notice the fact that two clusters can not come close to each other 
because, as just mentioned, of the inter-cluster Pauli repulsion, which implies that two clusters in the intrinsic state 
(the prolate THSR wave function) are effectively localized in space.  Thus, the prolate THSR wave function has the 
parity-violating deformation of localized $^{16}$O + $\alpha$ clustering. We can say that dynamics prefers 
nonlocalized clustering  but kinematics makes the system look like localized clustering. Of course, this localization is most pronounced in the necessarily strongly prolate two cluster systems. In systems with low density $\alpha$ clusters in number more than two have more space to move independently and are, therefore, less localized in spherical containers.

\begin{figure}[htbp]
\begin{center}
\includegraphics[scale=0.36]{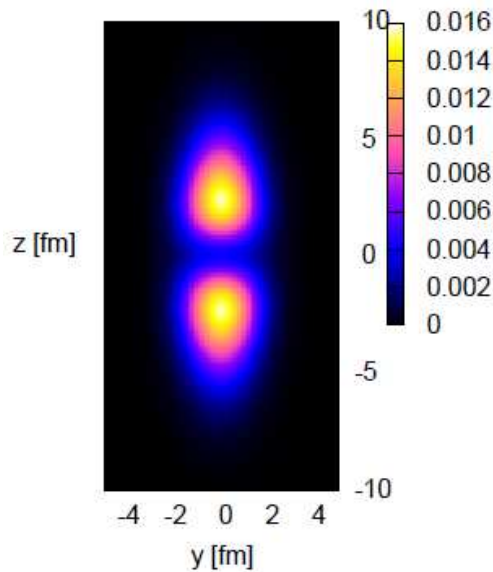}
\caption{Density distribution of a 2$\alpha$ prolate THSR wave function with $(\beta_x, \beta_y, \beta_z)$ = 
(1.78 fm, 1.78 fm, 7.85 fm). \label{fig:4}}
\end{center}
\end{figure}

The effective localization of clusters in the prolate THSR wave function of the two-cluster system is clearly seen in 
the density distribution of the prolate THSR wave function.   In Fig.~\ref{fig:4} we show the density 
distribution of a 2$\alpha$ prolate THSR wave function with $(\beta_x, \beta_y, \beta_z)$ = 
(1.78 fm, 1.78 fm, 7.85 fm).  Since the THSR wave function before the antisymmetrization operation is 
obviously composed of nonlocalized clusters, it is evident that the clear spatial localization of clusters shown 
in this figure is attributed to the inter-cluster Pauli principle. 

\begin{figure}[htbp]
\begin{center}
\includegraphics[scale=0.36]{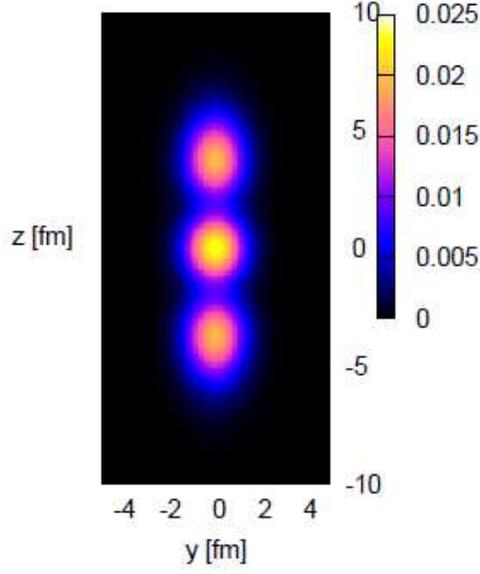}
\caption{Density distribution of a 3$\alpha$ THSR wave function with strong prolate deformation with 
$(\beta_x, \beta_y, \beta_z)$ = (0.01 fm, 0.01 fm, 5.1 fm). This figure is taken from Ref.~\cite{Su13}.\label{fig:5}}
\end{center}
\end{figure}

Recently it has been reported~\cite{Su13} that the density distribution of a 3$\alpha$ THSR wave function with 
strong prolate deformation with $(\beta_x, \beta_y, \beta_z)$ = (0.01 fm, 0.01 fm, 5.1 fm) shows clear spatial 
localization of three $\alpha$ clusters aligned linearly, which is displayed in Fig.~\ref{fig:5}.  It is to be 
noted that because of the almost zero values of $\beta_x=\beta_y$, three $\alpha$ clusters are not allowed 
to expand into the $x$ and $y$ directions, which means that three $\alpha$ clusters are only allowed to make 
one-dimensional motion along the $z$ direction.  Therefore the inter-cluster Pauli principle acts only along 
the $z$ direction, which is the reason of the spatial localization of the three $\alpha$ clusters.   In Ref.~\cite{Su13} it is reported that the $\alpha$ linear-chain Brink-GCM wave function is almost 
100$\%$ equivalent to a single 3$\alpha$ THSR wave function with strong prolate deformation which is just 
the $\alpha$ THSR wave function in Fig.~\ref{fig:5} having $(\beta_x, \beta_y, \beta_z)$ = 
(0.01 fm, 0.01 fm, 5.1 fm).   This 3$\alpha$ THSR wave function may be called a one-dimensional 
container-model wave function or a one-dimensional $\alpha$ particle condensate. Macroscopic boson condensates with inpenetrable (hard core) bosons are known under the name of ``Girardeau-Tonks" gases \cite{lieb2005}. In such cases the bosons behave like fermions. It may be an interesting study for the future in how much such a picture also is born out in linear-chain states of $\alpha$ particles.

\begin{figure}[htbp]
\begin{center}
\includegraphics[scale=0.36]{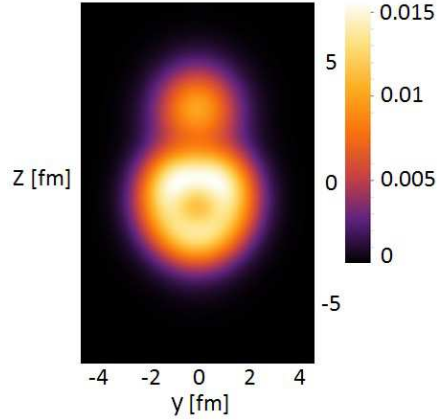}
\caption{Density distribution of the $^{16}$O + $\alpha$ hybrid-Brink-THSR wave function with $S_z$ = 0.6 fm and 
$(\beta_x, \beta_y, \beta_z)$ = (0.9 fm, 0.9 fm, 2.5 fm).\label{fig:6}}
\end{center}
\end{figure}

Let us now investigate the effectively spatial localization of the $^{16}$O and $\alpha$ clusters in the prolate THSR wave 
function of $^{16}$O + $\alpha$ system.   For this purpose we first notice the fact that the THSR wave function 
is a state of good parity.  Therefore a pure THSR wave function is not suitable for expressing a parity-breaking 
density distribution of the $^{16}$O-$\alpha$ clustering.  However, as we will see below, if we use a hybrid-Brink-THSR 
wave function with small $S_z$ parameter, the density distribution of this hybrid-Brink-THSR 
wave function which is quite close to a prolate THSR wave function shows clearly the effectively spatial localization of 
$^{16}$O and $\alpha$ clusters.  In Fig.~\ref{fig:6} we show the density distribution of the hybrid-Brink-THSR 
wave function with $S_z$ = 0.6 fm and $(\beta_x, \beta_y, \beta_z)$ = (0.9 fm, 0.9 fm, 2.5 fm).  We observe in 
this figure that, in spite of the small value of $S_z$ = 0.6 fm, the inter-cluster distance between $^{16}$O and 
$\alpha$ is about 3.6 fm.  Namely the large inter-cluster distance of about 3.6 fm between $^{16}$O and $\alpha$ is not 
due to the parameter $S_z$ but due to the effective spatial localization of $^{16}$O and $\alpha$ clusters in  
the prolate THSR wave function with $(\beta_x, \beta_y, \beta_z)$ = (0.9 fm, 0.9 fm, 2.5 fm) which is just the 
wave function reduced from the hybrid-Brink-THSR wave function by letting go $S_z$ to zero. Please also notice that for small values of $S_z$ the energies in Fig.~\ref{nonlocalized} are practically degenerate. Also it should be noted that Fig.~\ref{fig:6} corresponds to the ground state of \nene\ and, thus, has a much higher average density than the one of ${}^{8}$Be. Therefore, the cluster structure is more compact.

In the above we discussed that, in two-cluster systems, cluster states generally have effective localization of 
clusters because of the inter-cluster Pauli repulsion.   However, in three or more cluster systems, the spatial 
arrangement of clusters are not necessarily geometrical, namely clusters can be nonlocalized, although  
the inter-cluster separations are non-zero simultaneously because of the inter-cluster Pauli repulsion.  
However, as is discussed in Ref.~\cite{Su13}, if a cluster state is forced to have strongly-prolate deformation, 
the state can have effective localization of clusters like in the case of 3$\alpha$ linear-chain structure 
of Fig.~\ref{fig:4}.    When the inter-cluster separations are large, the spatial arrangement of clusters can 
be non-rigid  and gas-like. More on the spatial behaviour of three (or more) $\alpha$ particle systems will be investigated in the future.

As is seen in the above discussions, the container picture of cluster dynamics has three important ingredients.  
The first is to regard the motion of clusters as being mutually independent and described by the nonlocalized 
lowest orbit of the self-consistent mean-field potential of clusters.   The second is the collective excitation 
of the system which is described by the Hill-Wheeler equation with respect to the size parameter(s) $B$ of the 
mean-field potential.  The third is the inter-cluster Pauli repulsion which, in the case of two-cluster systems,  
is the origin of the molecular structure of clusters and which, in cases of more $\alpha$ clusters, like in the Hoyle state of \cc, leads to $\alpha-\alpha$ scattering processes which somewhat depopulate the $\alpha$  condensate.

\section{Summary, discussions, and outlook  \label{sec6}}

In this paper we first discussed the hybrid-Brink-THSR wave function introduced in Ref.~\cite{Zh13}. 
The energy curve with this new-type of wave function revealed that the traditional understanding is incorrect, namely that the 
localized-cluster picture is strongly supported by the energy curve with the Brink wave function which gives the minimum point 
at a non-zero value of the inter-cluster distance parameter.  The relative wave function of the Brink wave 
function is a Gaussian wave packet with  fixed size parameter $S_z$, 
\begin{eqnarray}
  \exp[-\frac{A_1A_2}{A_1+A_2}\frac{1}{2b^2}(\Vec{r}-S_z\Vec{e}_z)^2],
\end{eqnarray}
for a two-cluster system with mass numbers $A_1$ and $A_2$ with $b$ standing for the usual H.O. 
size parameter in the ground state Slater determinant, while that of the hybrid-Brink-THSR wave function is a Gaussian wave packet with variable size 
parameter, 
\begin{eqnarray}
  \exp[-\frac{A_1A_2}{A_1+A_2}\frac{1}{2B^2}(\Vec{r}-S_z\Vec{e}_z)^2], \quad B^2 = b^2 + 2\beta^2.
\end{eqnarray}
The minimum point of the energy curve with the hybrid-Brink-THSR wave function has a non-zero $S_z$ value when 
$B = b$ which is the limit case of the Brink wave function, but as $B$ becomes larger the $S_z$ value of 
the energy-minimum point becomes smaller and reaches $S_z = 0$ which is the limit case of the THSR wave function. 
Namely if we allow the size parameter of relative wave function of the Brink wave function to take an arbitrary value, 
the minimum point of the energy curve is no more at non-zero inter-cluster distance parameter but at zero inter-cluster 
distance parameter.  The energy-minimum point at the limit of the THSR wave function is very different from that 
at the limit of the Brink wave function in their characters, because the THSR wave function at the energy-minimum point 
is almost 100$\%$ equivalent to the full solution of RGM while the Brink wave function at the energy-minimum point 
is only the main component of the full solution of RGM.  In two-cluster systems, the almost 100$\%$ equivalence of 
the full solution of the RGM to a single THSR wave function has been confirmed in the ground-state band states of $^8$Be and 
in the inversion-doublet band states of $^{20}$Ne.  
In the case of three-cluster systems, we know that about 93$\%$ equivalence of the full solution of the 
$3\alpha$ RGM to a single THSR wave function has been found in the {\it ground state} of $^{12}$C while almost 100$\%$ 
equivalence of the full solution of the $3\alpha$ RGM to a single THSR wave function has been found in the 
{\it Hoyle state} of $^{12}$C.  It will be important to check whether this kind of high-percentage equivalence of the 
full solution of the RGM to a single THSR wave function is true or not in general three-cluster and also 
more-than-three-cluster systems. 

The THSR wave function which was originally devised for the description of cluster-gas states has proved to be 
able also to describe well non-gas cluster states including ground states with more or less pronounced cluster structure. 
The ground state of $^{12}$C and the ground-state band states of $^{20}$Ne are largely of shell-model character.  The 
good reproduction of these states by THSR wave functions means that the THSR wave function can also well express 
shell model characters.  This point is assured by the property of the THSR wave function that in the limit 
of $B \to b$ the THSR wave function is equal to some important shell-model wave function. In the Appendix B, we discuss two examples concerning the \oo\ + $\alpha$ system and the 3$\alpha$ system in the ground state of \cc.

    We also elaborated on the container picture of the cluster dynamics. It is firstly described by the Hill-Wheeler 
equation of the size parameter $B$ of the THSR wave function which, in the case of the spherical THSR wave 
function, is written as 
\begin{eqnarray}
   \sum_j \langle \Phi_L (B_i)| (H - E) |\Phi_L (B_j) \rangle f(B_j) = 0.    \label{eq:hillwheeler}
\end{eqnarray}
Here the integration over $B$ in the Hill-Wheeler equation is expressed by the summation over the discrete 
values of $B$.  In the case of two-cluster systems where the THSR wave function $\Phi_L (B)$ is written as 
\begin{eqnarray}
   \Phi_L (B_k) = {\cal A} \{ r^L \exp(-\gamma_k r^2) Y_{L0}(\widehat{r}) \phi(C_1) \phi(C_2)\}, \quad 
     \gamma_k = \frac{A_1A_2}{A_1+A_2} \frac{1}{2B_k^2},    
\end{eqnarray}  
this  Hill-Wheeler equation is equivalent to the RGM equation.  It is because this Hill-Wheeler equation can 
be rewitten as 
\begin{eqnarray}
  && \langle Y_{L0}(\widehat{r}) \phi(C_1) \phi(C_2) | (H - E) |
     {\cal A} \{ \chi_L(r) Y_{L0}(\widehat{r}) \phi(C_1) \phi(C_2)\} \rangle  = 0,     \\
  && \chi_L(r) = \sum_j f(B_j) r^L \exp(-\gamma_j r^2).   
\end{eqnarray} 
From this equivalence we can conclude that we can solve the scattering problem with the Hill-Wheeler equation 
of the THSR wave function. In the cases of the ground-state band of ${}^{8}$Be and the inversion-doublet band states of \nene, the obtained THSR-GCM wave functions were found to be almost 100$\%$ equivalent to {\it single} THSR wave functions. On the other hand it was pointed out that, since the THSR-GCM wave functions are equal to the Brink-GCM ones, the latter are also equivalent to single THSR wave functions. This fact naturally implies that the structure of the THSR wave function captures very well the clustering dynamics.

The THSR wave function can describe in a unified and natural manner three kinds of states, the ground state, the ordinary cluster state, and the alpha-condensate state. The Brink-GCM wave function, that is the superposition of localized Brink wave function, sometimes demands large efforts to describe some excited states of cluster nature as, e.g., the Hoyle state of \cc. The Hill-Wheeler equation of Eq.~(\ref{eq:hillwheeler}) was solved for the systems of 
3$\alpha$'s~\cite{To01,Fu03,Fu05} and 4$\alpha$'s~\cite{To01,Fu10}.  In the $3\alpha$ system the THSR wave functions were shown to successfully reproduce the ground state and the Hoyle state with a $3\alpha$ condensate-like structure, that is with a large value of $B$-parameter underlying the container picture. It is to be recalled that even in the $3\alpha$ case the THSR-GCM wave function of the ground state has 93$\%$ squared overlap with a single THSR wave function and that the THSR-GCM wave function of the Hoyle state has 99$\%$ squared overlap with a single THSR wave function~\cite{Fu09}. In fact, these squared overlaps are expected to grow to nearly 100$\%$ if $2\alpha$ correlations are included to the container picture. It also is to be noted that even in the case of $3\alpha$'s the THSR-GCM wave functions are almost 100$\%$ equivalent to Brink-GCM ones.
We also succeeded to calculate the $J^\pi = 2^+$ states and to reproduce the ground-state band $2^+$ state and the $3\alpha$ gas-like $2_2^+$ state.   In the $4\alpha$ system we succeeded to reproduce, for the spin $J^\pi = 0^+$, the ground state and the 
$0_6^+$ state which is the Hoyle-analogue state with a $4\alpha$ condensate-like structure. 
However, unlike for the $3\alpha$ system, in the $4\alpha$ system, the cluster states lying between the ground state 
and the $\alpha$ condensate-like $0_6^+$ state could not be fully reproduced with the Hill-Wheeler equation of 
Eq.~(\ref{eq:hillwheeler}).  Namely instead of the observed four cluster states between the ground state and the 
$0_6^+$ state (which were nicely reproduced by the $4\alpha$ OCM of Ref.~\cite{Fu08a}), the Hill-Wheeler equation of Eq.~(\ref{eq:hillwheeler}) could give us only two states.  The reason of this insufficiency is because of the 
variety of the observed cluster states in $^{16}$O for which the use of only one collective coordinate $B$ 
is too simple and unsatisfactory.  The result of the $4\alpha$ OCM calculation of Ref.~\cite{Fu08a} tells us 
that the dominant structures of $0^+_2$, $0^+_3$, $0^+_4$, and $0^+_5$ states are $^{12}$C$(0^+_1)$ + 
$\alpha$ ($S$ wave), $^{12}$C$(2^+_1)$ + $\alpha$ ($D$ wave), $^{12}$C$(0^+_1)$ + $\alpha$ 
($S$ wave with higher nodal number), and $^{12}$C$(1^-_1)$ + $\alpha$ ($P$ wave), respectively. 
One possible way to cope with this variety of cluster structures is to use two kinds of size parameters $\Vec{B}_1$ 
and $\Vec{B}_2$ by extending the THSR wave function;  
\begin{eqnarray}
  {\cal A} \left\{ \exp\left(-\frac{3}{2B_2^2} r^2\right) \Phi_{3\alpha}(\Vec{B}_1) \phi(\alpha_4) \right\}, 
  \quad \Vec{r} = \Vec{X}_4 - \frac{1}{3}\sum_{j=1}^3 \Vec{X}_j,   \label{eq:c12-alpha}
\end{eqnarray} 
where $\Phi_{3\alpha}(\Vec{B}_1)$ is the THSR wave function of three $\alpha$ clusters, $\alpha_1 \sim \alpha_3$, 
with the deformed size parameter $\Vec{B}_1$ with $3\alpha$ center-of-mass coordinate removed.  This extended THSR wave function is similar to the RGM-type wave function of two clusters, $\Phi_{3\alpha}(\Vec{B}_1)$ and 
$\phi(\alpha_4)$, with the relative wave function $\exp(-\gamma r^2)$ with $\gamma = 3/(2B_2^2)$.  
Therefore the superposition of this wave function over $B_2$ allows us to describe scattering states in various 
channels of $^{12}$C + $\alpha$. This will also be a very interesting study for the near future.

The original THSR wave function describes only positive-parity states.  Therefore in Ref.~\cite{Zh13} where the 
negative-parity partner band of the inversion doublet bands had to be studied, a prescription was proposed 
for constructing negative-parity wave functions as a natural extension of the original THSR wave function. 
This prescription is for two-cluster wave functions.   In Ref.~\cite{Zh13} and also in this paper we call the 
negative-parity wave functions constructed by this prescription simply THSR wave functions. As a starting point,
we used a hybrid-Brink-THSR wave function with a non-zero inter-cluster separation parameter $S_z$. 
From this wave function we project out the negative-parity wave function and then normalize it.  After 
normalization we can take safely the limit of $S_z \to$ 0, and this limit wave function is just the 
negative-parity THSR wave function.  This prescription to construct the negative-parity THSR wave function 
in two-cluster systems can be generalized to the systems with three and more clusters.  For instance, 
in the case of $4\alpha$ system, we extend the $3\alpha$ - $\alpha$ THSR wave function of Eq.~(\ref{eq:c12-alpha}) 
into the hybrid-Brink-THSR-type wave function, 
\begin{eqnarray}
  {\cal A} \left\{ \exp\left(-\frac{3}{2B_2^2} (\Vec{r} - \Vec{S})^2\right) \Phi_{3\alpha}(\Vec{B}_1) 
  \phi(\alpha_4) \right\}. 
\end{eqnarray} 
From this wave function we project out the negative-parity wave function and then normalize it.  After 
normalization we can take safely the limit of $|\Vec{S}| \to$ 0, and this limit wave function is just the 
negative-parity THSR wave function which we intend to construct.  

The container picture of cluster dynamics has three important ingredients.  The first is the mutually 
independent nonlocalized motion of clusters occupying the lowest orbit of the self-consistent mean-field 
potential of clusters.   The second is the collective motion with respect to the size parameter(s) $B$ 
of the mean-field potential which is described by the Hill-Wheeler equation of Eq.~(\ref{eq:hillwheeler}). 
The third is the inter-cluster Pauli repulsion due to the Fermi statistics of the nucleons which constitute     
the clusters. In two-cluster systems, this Pauli repulsion makes the two clusters locate at some distance from each other, which gives rise, effectively, to localized clustering in two-cluster systems, in spirit similar to the phenomenological excluded volume prescription.  This is the reason why the 
cluster states in two-cluster systems are always molecular states with spatial localization of 
clusters.  However, in the systems of three or more clusters, although the inter-cluster separations in all 
pairs of clusters are non-zero simultaneously, it does not necessarily mean in general the formation of some  
localized arrangement of clusters.  In spite of this general situation, we know that there have been reports 
of localized cluster structure in systems of three or more clusters.   For example, the existence of an 
excited $0^+$ state with somewhat bent linear-chain configuration of $3\alpha$'s has been predicted by the 
AMD calculation of Ref.~\cite{En98,En07} and also by the FMD calculation of Ref.~\cite{Ne03,Ch07}. 
The formation of this quasi-linear $3\alpha$ state is argued to be due to the orthogonality of this state 
to the ground state and the Hoyle state of $^{12}$C with the AMD study~\cite{Su13a} and with the
THSR wave function~\cite{Fu13}.   Another example is the study of the $4\alpha$ linear-chain state with high 
angular momentum of Ref.~\cite{Ic11}.   Here, the formation of the linear-chain structure is considered 
dominantly due to the effect of the centrifugal force of high-spin rotation. As a final remark, let us say that in this and the preceding work \cite{Zh12, Zh13}, we have extended the THSR wave function to negative-parity states, we here further sketched how in incorporating more size parameters into the THSR wave function a much richer flexibility may be reached, adapted for the description of more complicated cluster configuration involving several clusters of different sizes together with their proper, possible internal cluster configurations as, e.g., described above for the $^{12}$C case.

\begin{acknowledgments}
This work is supported by the National Natural Science Foundation of China (Nos. 11035001, 10975072, 10735010, 11375086, 11175085, 11235001, and 11120101005), by the 973 Program of China (No. 2010CB327803 and No. 2013CB834400), by CAS Knowledge Innovation Project No. KJCX2-SW-N02, by Research Fund of Doctoral Point (RFDP) (No. 20100091110028), and by the Project Funded by the Priority Academic Program Development of Jiangsu Higher Education Institutions (PAPD).
\end{acknowledgments}

\clearpage
\appendix{\bf Appendix A: Analytic formula of the quadrupole moment by two-cluster wave function}
\vspace{0.3cm}

We derive an analytic formula of the expectation value of the quadrupole-moment operator by 
two-cluster RGM wave function.   We treat the case where both clusters of the system are SU(3)-scalar nuclei, 
such as $\alpha$, $^{16}$O, and $0s$-shell nuclei like $d$, $t$, $^3$He.   For simplicity we consider only the case 
where the isospin of the total system is zero. The expectation value $Q(L)$ of the quadrupole-moment operator is 
expressed as follows
\begin{eqnarray}
  && Q(L) = \langle \Psi | \sum_i \frac{1}{2} (1 + (\tau_z)_i) Q_{20}(i) | \Psi \rangle 
     = \langle \Psi | \frac{1}{2} \sum_i Q_{20}(i) | \Psi \rangle,    \label{eq:qmom}    \\
  && \Psi = \sqrt{\frac{A_1!A_2!}{A!}} {\cal A} \left\{ \chi_L(r)Y_{LL}(\widehat{r}) \phi(C_1) \phi(C_2)\right\}, 
     \quad \langle \Psi | \Psi \rangle = 1, \quad \chi_L(r) = {\rm arbitrary}, \\
  && Q_{20}(i) = Q_{20}(\Vec{r}_i - \Vec{X}_G), \quad Q_{20}(\Vec{R}) = 3R_z^2 - R^2 =  
    \sqrt{\frac{16\pi}{5}} R^2 Y_{20}(\widehat{R}), 
\end{eqnarray}
where $\Vec{X}_G$ is the total center-of-mass coordinate.  In calculating $Q(L)$, we use the following identity 
relation, 
\begin{eqnarray}
  && \sum_i Q_{20}(i) =  Q_{20}(C_1) + Q_{20}(C_2) + \frac{A_1 A_2}{A} Q_r, \\
  &&  Q_{20}(C_k) =  \sum_{i \in C_k} Q_{20} (\Vec{r}_i - \Vec{X}_{Gk}),  \\ 
  &&  Q_r = Q_{20}(\Vec{r}) = \sqrt{\frac{16\pi}{5}} r^2 Y_{20}(\widehat{r}), 
\end{eqnarray}
where $\Vec{X}_{Gk}$ stands for the center-of-mass coordinate vector of the $k$-th cluster $C_k$. 

Now we discuss the calculation of the matrix element $q_L(N, N')$ of the quadrupole-moment operator by the H.O. 
(harmonic oscillator) basis wave function of RGM
\begin{eqnarray}
  && q_L(N, N') =     \nonumber \\ 
  && \langle R_{NL}(r)Y_{LL}(\widehat{r}) \phi(C_1) \phi(C_2)| \frac{1}{2} \sum_i Q_{20}(i)| 
       {\cal A} \left\{ R_{N'L}(r)Y_{LL}(\widehat{r}) \phi(C_1) \phi(C_2) \right\} \rangle,     
\end{eqnarray}
where $R_{NL}(r)$ is the H.O. radial function with $N$ standing for $N = 2n + L$ and $n$ standing for the 
number of nodal points.   When $N > N'$, we operate $\sum_i Q_{20}(i)$ on the bra side and we get 
\begin{eqnarray}
  && q_L(N, N') =  \frac{1}{2} \langle (Q_{20}(C_1) + Q_{20}(C_2) + \frac{A_1 A_2}{A} Q_r)
     R_{NL}(r)Y_{LL}(\widehat{r}) \phi(C_1) \phi(C_2)     \nonumber   \\
  && \hspace{1cm}  | {\cal A} \left\{ R_{N'L}(r)Y_{LL}(\widehat{r}) \phi(C_1) \phi(C_2) \right\} \rangle    \\
  && \hspace{0.5cm} = \delta_{N,N'+2}\ \frac{A_1 A_2}{2A} \langle  R_{N'L}(r)Y_{LL}(\widehat{r})| Q_r| 
     R_{NL}(r)Y_{LL}(\widehat{r}) \rangle \mu_{N'L}.   \label{eq:qNgtNp}
\end{eqnarray}
Here $\mu_{NL}$ is defined as 
\begin{eqnarray}
  \mu_{NL} = \langle R_{NL}(r)Y_{LL}(\widehat{r}) \phi(C_1) \phi(C_2) | {\cal A} \left\{ R_{NL}(r)Y_{LL}(\widehat{r}) 
  \phi(C_1) \phi(C_2) \right\} \rangle. 
\end{eqnarray}
It is known that $\mu_{NL}$ depends on $N$ but not on $L$ in the system composed of two SU(3)-scalar 
clusters~\cite{Ho74,Ho77}.  In obtaining Eq.~(\ref{eq:qNgtNp}) we used the fact that the number of H.O. quanta 
of $Q_{20}(C_k) \phi(C_k)$ is equal to or larger than that of $\phi(C_k)$ which is the reason of no contribution 
from the operator $Q_{20}(C_k)$.  We also used the fact that in the H.O. expansion of $Q_r R_{NL}(r)$, 
\begin{eqnarray}
  && Q_r R_{NL}(r) =  R_{N-2,L}(r) \langle R_{N-2,L}(r)|Q_r| R_{NL}(r)\rangle + 
     R_{NL}(r) \langle R_{NL}(r)|Q_r| R_{NL}(r)\rangle   \nonumber \\
  && \hspace{1cm} + R_{N+2,L}(r) \langle R_{N+2,L}(r)|Q_r| R_{NL}(r)\rangle, 
\end{eqnarray}
only the term $R_{N-2,L}(r) \langle R_{N-2,L}(r)|Q_r| R_{NL}(r)\rangle$ can survive because of the conservation 
of the number of the H.O. quanta between bra and ket RGM basis states.

When $N < N'$, we operate $\sum_i Q_{20}(i)$ on the ket side and we get 
\begin{eqnarray}
  && q_L(N, N') =  \frac{1}{2} \langle R_{NL}(r)Y_{LL}(\widehat{r}) \phi(C_1) \phi(C_2)      \nonumber   \\
  && \hspace{1cm}  | {\cal A} \left\{ (Q_{20}(C_1) + Q_{20}(C_2) + \frac{A_1 A_2}{A} Q_r) 
     R_{N'L}(r)Y_{LL}(\widehat{r}) \phi(C_1) \phi(C_2) \right\} \rangle    \\
  && \hspace{0.5cm} = \delta_{N+2,N'}\ \frac{A_1 A_2}{2A} \langle  R_{NL}(r)Y_{LL}(\widehat{r})| Q_r| 
     R_{N'L}(r)Y_{LL}(\widehat{r}) \rangle \mu_{NL}.     \label{eq:qNgtNm}
\end{eqnarray}
Finally, when $N = N'$, we get 
\begin{eqnarray}
  && q_L(N, N') =  \frac{1}{2} \langle R_{NL}(r)Y_{LL}(\widehat{r}) \phi(C_1) \phi(C_2)     \nonumber  \\
  && \hspace{1cm} |{\cal A} \left\{ (Q_{20}(C_1) + Q_{20}(C_2) + \frac{A_1 A_2}{A} Q_r)
     R_{NL}(r)Y_{LL}(\widehat{r}) \phi(C_1) \phi(C_2)\right\} \rangle   \\
  && \hspace{0.5cm} =  \frac{A_1 A_2}{2A} \langle  R_{NL}(r)Y_{LL}(\widehat{r})| Q_r| 
     R_{NL}(r)Y_{LL}(\widehat{r}) \rangle \mu_{NL}.   \label{eq:qNgtNz}
\end{eqnarray}
Here we used the fact that the cluster wave function $\phi(C_k)$ is the only one  wave function 
which has the smallest number of the total H.O. quanta in the mass-number $A_k$ system.  
The closed-shell wave functions of $\phi(\alpha)$ and $\phi(^{16}{\rm O})$ and also the wave functions of 
$0s$-shell nuclei have this property.  In order to fulfill the conservation of the number of the H.O. quanta 
between bra and ket RGM basis states, in the expansion of $Q_{20}(C_k) \phi(C_k)$ 
\begin{eqnarray}
  Q_{20}(C_k) \phi(C_k) = \sum_j \Psi_j \langle \Psi_j |Q_{20}(C_k) \phi(C_k) \rangle,
\end{eqnarray}
only the expansion state $\Psi_{j=j_0}$ having the same number of the H.O. quanta as $\phi(C_k)$ can make non-zero 
contribution.  But as we mentioned above, such $\Psi_{j=j_0}$ is nothing but $\phi(C_k)$ itself.  Therefore 
because of $\langle \phi(C_k)| Q_{20}(C_k)| \phi(C_k) \rangle$ =  0, $Q_{20}(C_k)$ makes no contribution, and 
we get the result of Eq.~(\ref{eq:qNgtNz}). 

Summarizing Eqs. (\ref{eq:qNgtNp}), (\ref{eq:qNgtNm}), and (\ref{eq:qNgtNz}), we obtain 
\begin{eqnarray}
  q_L(N, N') &=& \frac{1}{2} \sqrt{\frac{16\pi}{5}} \langle Y_{LL}(\widehat{r})| Y_{20}(\widehat{r})| 
     Y_{LL}(\widehat{r}) \rangle \ r_L(N, N'),  \\
  r_L(N, N') &=& \left\{ \delta_{N,N'+2}\  \mu_{N'L} + \delta_{N+2,N'}\ \mu_{NL} + \delta_{N,N'}\ \mu_{NL} \right\} 
     \nonumber \\
  && \times  \frac{A_1 A_2}{A} \langle R_{NL}(r)| r^2| R_{N'L}(r) \rangle.        
\end{eqnarray}

The quantity $r_L(N, N')$ is intimately related to the matrix element $R_L(N, N')$ of the square radius operator 
$\sum_i (\Vec{r}_i - \Vec{X}_G)^2$ by the H.O. basis wave function of RGM
\begin{eqnarray}
  && R_L(N, N') =     \nonumber \\ 
  && \langle R_{NL}(r)Y_{LL}(\widehat{r}) \phi(C_1) \phi(C_2)| \sum_i (\Vec{r}_i - \Vec{X}_G)^2| 
       {\cal A} \left\{ R_{N'L}(r)Y_{LL}(\widehat{r}) \phi(C_1) \phi(C_2) \right\} \rangle,     
\end{eqnarray}
In order to calculate $R_L(N, N')$, we express the operator $\sum_i (\Vec{r}_i - \Vec{X}_G)^2$ as follows 
\begin{eqnarray}
  && \sum_i (\Vec{r}_i - \Vec{X}_G)^2 = R^2(C_1) + R^2(C_2) + \frac{A_1 A_2}{A} r^2,  \\
  && R^2(C_k) = \sum_{i \in C_k} (\Vec{r}_i - \Vec{X}_{Gk})^2, \ (k=1,2). 
\end{eqnarray}
By using this expression of the square radius operator, we can make the calculation of $R_L(N, N')$ just 
in the same way as that of $q_L(N, N')$.   For $N > N'$, 
\begin{eqnarray}
  && R_L(N, N') = \langle  (R^2(C_1) + R^2(C_2) + \frac{A_1 A_2}{A} r^2)
     R_{NL}(r)Y_{LL}(\widehat{r}) \phi(C_1) \phi(C_2)      \nonumber  \\
  && \hspace{1cm}  |{\cal A} \left\{ R_{N'L}(r)Y_{LL}(\widehat{r}) \phi(C_1) \phi(C_2) \right\}  \rangle    \\
  && \hspace{0.5cm} = \delta_{N,N'+2}\ \frac{A_1 A_2}{A} \langle  R_{N'L}(r)Y_{LL}(\widehat{r})| r^2| 
     R_{NL}(r)Y_{LL}(\widehat{r}) \rangle \mu_{N'L}.
\end{eqnarray}
For $N < N'$, 
\begin{eqnarray}
  && R_L(N, N') = \langle R_{NL}(r)Y_{LL}(\widehat{r}) \phi(C_1) \phi(C_2)    \nonumber    \\
  && \hspace{1cm} |{\cal A} \left\{ (R^2(C_1) + R^2(C_2) + \frac{A_1 A_2}{A} r^2) 
     R_{N'L}(r)Y_{LL}(\widehat{r}) \phi(C_1) \phi(C_2)\right\} \rangle    \\
  && \hspace{0.5cm} = \delta_{N+2,N'}\ \frac{A_1 A_2}{A} \langle  R_{NL}(r)Y_{LL}(\widehat{r})| r^2| 
     R_{N'L}(r)Y_{LL}(\widehat{r}) \rangle \mu_{NL}.
\end{eqnarray}
For $N = N'$, 
\begin{eqnarray}
&& R_L(N, N') = \langle R_{NL}(r)Y_{LL}(\widehat{r}) \phi(C_1) \phi(C_2)    \nonumber    \\
  && \hspace{1cm} |{\cal A} \left\{ (R^2(C_1) + R^2(C_2) + \frac{A_1 A_2}{A} r^2)
     R_{NL}(r)Y_{LL}(\widehat{r}) \phi(C_1) \phi(C_2)\right\} \rangle    \\
  && \hspace{0.5cm} =  \left\{ \langle R^2(C_1) \rangle + \langle R^2(C_2) \rangle + 
     \frac{A_1 A_2}{A} \langle  R_{NL}(r)Y_{LL}(\widehat{r})| r^2| 
     R_{NL}(r)Y_{LL}(\widehat{r}) \rangle \right\} \mu_{NL},  \\
  && \langle R^2(C_k) \rangle = \langle \phi(C_k)| R^2(C_k)| \phi(C_k) \rangle, \ (k=1, 2). 
\end{eqnarray}
From these results we have 
\begin{eqnarray}
  R_L(N, N') = r_L(N, N') + \left(\langle R^2(C_1) \rangle + \langle R^2(C_2) \rangle \right)\ \mu_{NL}\ \delta_{N,N'}. 
\end{eqnarray}
The relation of $q_L(N, N')$ and $R_L(N, N')$ is 
\begin{eqnarray}
  && q_L(N, N') = \frac{1}{2} \sqrt{\frac{16\pi}{5}} \langle Y_{LL}(\widehat{r})| Y_{20}(\widehat{r})| 
     Y_{LL}(\widehat{r}) \rangle \ r_L(N, N')  \\
  && = \frac{1}{2} \sqrt{\frac{16\pi}{5}} \langle Y_{LL}(\widehat{r})| Y_{20}(\widehat{r})| 
     Y_{LL}(\widehat{r}) \rangle \nonumber \\ 
  && \hspace{1cm} \times \left\{ R_L(N, N') 
     - (\langle R^2(C_1) \rangle + \langle R^2(C_2) \rangle) \mu_{NL} \delta_{N,N'} \right\}.  
\end{eqnarray}

The calculation of $Q(L)$ of Eq.~(\ref{eq:qmom}) is now made as follows.  First we expand the relative wave function 
$\chi_L(r)$ by H.O. functions 
\begin{eqnarray}
  && \chi_L(r) = \sum_N C_N  R_{NL}(r), \\
  && 1 = \langle \Psi | \Psi \rangle = \langle \chi_L(r)Y_{LL}(\widehat{r}) \phi(C_1) \phi(C_2) | 
     {\cal A} \left\{ \chi_L(r)Y_{LL}(\widehat{r}) \phi(C_1) \phi(C_2)\right\} \rangle  \\
  && \hspace{1cm} = \sum_N (C_N)^2 \mu_{NL}.    
\end{eqnarray}
Then the Q-moment of $\Psi$ is calculated as 
\begin{eqnarray}
  && \langle \Psi | \frac{1}{2} \sum_i Q_{20}(i) | \Psi \rangle \nonumber \\
  && = \langle \chi_L(r)Y_{LL}(\widehat{r}) \phi(C_1) \phi(C_2) | \frac{1}{2} \sum_i Q_{20}(i) | 
     {\cal A} \left\{ \chi_L(r)Y_{LL}(\widehat{r}) \phi(C_1) \phi(C_2)\right\} \rangle  \\
  && = \sum_{N,N'} C_N C_{N'} q_L(N,N') \\
  && = \frac{1}{2} \sqrt{\frac{16\pi}{5}} \langle Y_{LL}(\widehat{r})| Y_{20}(\widehat{r})| 
     Y_{LL}(\widehat{r}) \rangle \nonumber \\ 
  &&  \hspace{0.5cm} \times \sum_{N,N'} C_N C_{N'}  \left\{ R_L(N, N') 
     - (\langle R^2(C_1) \rangle + \langle R^2(C_2) \rangle) \mu_{NL} \delta_{N,N'} \right\} \\
  && = \frac{1}{2} \sqrt{\frac{16\pi}{5}} \langle Y_{LL}(\widehat{r})| Y_{20}(\widehat{r})| 
     Y_{LL}(\widehat{r}) \rangle \nonumber \\ 
  &&  \hspace{0.5cm} \times \left\{ \langle \Psi | \sum_i (\Vec{r}_i - \Vec{X}_G)^2 | \Psi \rangle 
     - (\langle R^2(C_1) \rangle + \langle R^2(C_2) \rangle)  \right\} \\
  && = \frac{1}{2} \sqrt{\frac{16\pi}{5}} \langle Y_{LL}(\widehat{r})| Y_{20}(\widehat{r})| 
     Y_{LL}(\widehat{r}) \rangle \frac{A_1 A_2}{A} \langle r^2 \rangle, \\
  && \frac{A_1 A_2}{A} \langle r^2 \rangle \equiv \langle \Psi | \sum_i (\Vec{r}_i - \Vec{X}_G)^2 | \Psi \rangle 
     - (\langle R^2(C_1) \rangle + \langle R^2(C_2) \rangle).        
\end{eqnarray}
Since we have 
\begin{eqnarray}
  && \langle Y_{LL}(\widehat{r})| Y_{20}(\widehat{r})|  Y_{LL}(\widehat{r}) \rangle 
     = \sqrt{\frac{5}{4\pi}} (L L 2 0| L L)\ (L 0 2 0| L 0),  \\
  && (L L 2 0| L L) = \sqrt{\frac{L(2L-1)}{(L+1)(2L+3)}}, \quad 
     (L 0 2 0| L 0) = - \sqrt{\frac{L(L+1)}{(2L-1)(2L+3)}},    \\ 
  && \frac{1}{2} \sqrt{\frac{16\pi}{5}} \langle Y_{LL}(\widehat{r})| Y_{20}(\widehat{r})| 
     Y_{LL}(\widehat{r}) \rangle = (L L 2 0| L L)\ (L 0 2 0| L 0) = - \frac{L}{2L+3},    
\end{eqnarray}     
we get the following final result
\begin{eqnarray}
   \langle \Psi | \frac{1}{2} \sum_i Q_{20}(i) | \Psi \rangle = - \frac{L}{2L+3} 
     \frac{A_1 A_2}{A} \langle r^2 \rangle.   
\end{eqnarray}
\vspace{1cm}

\appendix{\bf Appendix B: The shell model limits of the THSR wave functions in describing the ground states of \cc\ and \nene\ }
\vspace{0.3cm}

As we know, the THSR wave function can describe not only the gas-like cluster states but also the cluster states with normal density, even some shell-model-like ground states. Here, we give a detailed explanation why the THSR wave function at the limit of $B \to b$ can describe well the shell-model-like states of \cc\ and \nene.   

In the $^{16}$O + $\alpha$ system, the THSR wave function shown in Ref.~\cite{Zh13},  
${\cal A} \{ r^L \exp(-\gamma r^2) Y_{LM}(\widehat{r}) \phi(^{16}{\rm O}) \phi(\alpha)\}$ with even $L$ 
has the following character
\begin{eqnarray}
 && \lim_{B \to b} N_L(B) {\cal A} \{ r^L \exp(-\gamma r^2) Y_{LM}(\widehat{r}) \phi(^{16}{\rm O}) \phi(\alpha)\}  
    \nonumber \\
 && \quad  = n_L {\cal A} \{ R_{8L}(r,\gamma_0) Y_{LM}(\widehat{r}) \phi(^{16}{\rm O}) \phi(\alpha)\} 
    \label{eq:ne20expand}     \\ 
 && \quad = \psi((0s)^4(0p)^{12}(0d1s)^4;[4] (\lambda,\mu)=(8,0),LM)\ \frac{1}{g(\Vec{X}_G,20\nu)}, \label{eq:ne20BB} \\
 && g(\Vec{X}_G,20\nu) = (\frac{20\nu}{\pi})^{-3/4} \exp(-20\nu X_G^2),  \\ 
 && \gamma = \frac{8}{5}\frac{1}{B^2}, \quad \gamma_0 = \frac{8}{5}\frac{1}{b^2} = \frac{16}{5} \nu, 
    \quad \nu = \frac{1}{2b^2},      
\end{eqnarray}
where $N_L(B)$ and $n_L$ are normalization constants.  $n_L$ is independent of $L$, actually~\cite{Ho77}. 
$R_{N=8,L}(r,\gamma_0)$ is the radial H.O. function with size parameter $\gamma_0$ with $N$ standing for the 
number of H.O. quanta, $N = 2n + L$.  The equality of Eq.~(\ref{eq:ne20BB}) is due to the Bayman-Bohr 
theorem~\cite{Ba58,Ho77}.   Eq.~(\ref{eq:ne20BB}) shows that the THSR wave function becomes, in the limit 
of $B \to b$, the most important $sd$-shell shell-model wave function having spatial symmetry [4] and SU(3) symmetry 
$(\lambda,\mu)=(8,0)$.  
For deriving Eq.~(\ref{eq:ne20expand}) the following formula is useful 
\begin{eqnarray}
  \exp(-\gamma r^2) = \left(\frac{2\gamma}{\pi}\right)^{-\frac{3}{4}} 
     \left(\frac{2\sqrt{\gamma_0 \gamma}}{\gamma_0 + \gamma}\right)^{\frac{3}{2}} \sum_{n=0}^\infty 
     \sqrt{\frac{(2n+1)!!}{(2n)!!}} \left(\frac{\gamma - \gamma_0}{\gamma + \gamma_0}\right)^n 
     R_{2n,0}(r,\gamma_0)Y_{00}({\widehat r}).  \label{eq:expform}
\end{eqnarray}
The function $r^LR_{2n,0}(r,\gamma_0)$ has the form of $P_{2n+L}(r)\exp(-\gamma_0 r^2)$ where $P_{2n+L}(r)$ is 
a polynomial of $r$ with the highest-power term $r^{2n+L}$.   When we expand $P_{2n+L}(r)\exp(-\gamma_0 r^2)$ by 
the radial H.O. function $R_{N',L}(r,\gamma_0)$ as $P_{2n+L}(r) \exp(-\gamma_0 r^2) = \sum_{N'=0}^{N_0} C_{N'} 
R_{N',L}(r,\gamma_0)$, the maximum power $N_0$ is $N_0 = 2n + L$.  Since $R_{N',L}(r,\gamma_0) Y_{LM}(\widehat{r})$ 
with $N' < 8$ is Pauli-forbidden, we obtain Eq.~(\ref{eq:ne20expand}) in the limit of $B \to b$. 

As another example, we explain below the limit of $B \to b$ of the $3\alpha$ THSR wave function, 
\begin{eqnarray}
  &&\lim_{B \to b} N_{3\alpha}(B) {\cal A} \left\{ \exp\left(-(\frac{1}{B^2} \xi_1^2 + 
      \frac{4}{3B^2} \xi_2^2)\right) \prod_{i=1}^3 \phi(\alpha_i) \right\}   \\
  &&\quad = n_{3\alpha} {\cal A}\left\{F_4(\Vec{\xi}_1,\Vec{\xi}_2) \prod_{i=1}^3\phi(\alpha_i)\right\}  
      \label{eq:f4c12}   \\
  &&\quad = |(0s)^4(0p)^8,[444](0,4) J=0 \rangle\ \frac{1}{g(\Vec{X}_G,12\nu)},    \label{eq:c12ground}  \\
  &&F_n(\Vec{\xi}_1,\Vec{\xi}_2) = \sum_{n_1+n_2=n}\sqrt{\frac{(2n_1+1)!!(2n_2+1)!!}{(2n_1)!!(2n_2)!!}} 
      R_{2n_1,0}(\xi_1,\frac{1}{b^2})R_{2n_2,0}(\xi_2,\frac{4}{3b^2})      \nonumber \\
  &&\hspace{3.6cm} \times [Y_0({\widehat \xi}_1)Y_0({\widehat \xi}_2)]_{J=0},  \\
  &&g(\Vec{X}_G,20\nu) = (\frac{12\nu}{\pi})^{-3/4} \exp(-12\nu X_G^2), 
\end{eqnarray}
where $\Vec{\xi}_1$ and $\Vec{\xi}_2$ are inter-$\alpha$ Jacobi coordinates, 
$\Vec{\xi}_1 = \Vec{X}_2 - \Vec{X}_1$ and $\Vec{\xi}_2 = \Vec{X}_3 - (\Vec{X}_1 + \Vec{X}_2)/2$. 
$N_{3\alpha}(B)$ and $n_{3\alpha}$ are normalization constants. $F_n(\Vec{\xi}_1,\Vec{\xi}_2)$ is noted to be 
an eigen state of the H.O. quanta having the eigen value $2n$. 
The equality of Eq.~(\ref{eq:c12ground}) is because there is only one state in $^{12}$C which has total number of 
H.O. quanta $N=8=N_{min}$ and spatial symmetry $[444]$.  The equality of Eq.~(\ref{eq:f4c12}) is obtained by using
\begin{eqnarray}
  \exp\left(-(\frac{1}{B^2} \xi_1^2 + \frac{4}{3B^2} \xi_2^2)\right) \propto 
  \sum_{n=0}^{\infty} \left(\frac{b^2 - B^2}{b^2 + B^2}\right)^n F_n(\Vec{\xi}_1,\Vec{\xi}_2), 
\end{eqnarray}
which is due to Eq.~(\ref{eq:expform}), and by noting that since the lowest number of the total number of H.O. 
quanta ($N_{min}$) in $^{12}$C is 8, the terms with $n < 4$ in the above summation over $n$ vanish.  
Eq.~(\ref{eq:c12ground}) is one of the important reasons why the THSR wave function gives good description of 
the ground state of $^{12}$C. \\

\end{document}